\documentclass[a4paper,12pt,dvips]{article}
\usepackage{graphicx}
\usepackage{psfrag}
\usepackage{subfigure}
\usepackage{amsmath,amstext,amsfonts,amsbsy,amssymb,amscd,bbm,epsfig}
\usepackage{times}
=-0.5cm \oddsidemargin=-0.5cm

\title{ Finite temperature and confinement along the extra dimensions
studied on a five-dimensional U(1) lattice gauge model}

\author{K.~Farakos\footnote{E-mail: kfarakos@central.ntua.gr} ~and
S.~Vrentzos\footnote{E-mail: vrentsps@central.ntua.gr}}
\begin{document}
\maketitle
\begin{center}
Physics Department, National Technical University
of Athens,
\\ Zografou Campus 15780, Greece
\end{center}

\begin{abstract}
\noindent In this paper we study the properties of the phase diagram
of a simple extra dimensional model on the lattice at finite
temperature. We consider the five-dimensional pure gauge abelian
model with anisotropic couplings which at zero temperature exhibits
a new interesting phase, the layer phase. This  phase is
characterized by a massless photon living on the four dimensional
subspace and confinement along the extra dimension. We show that, as
long as the temperature takes a non zero value the aforementioned
layer phase disappears. It would be equivalent to assume that at
finite temperature the higher-dimensional lattice model loses any
feature of the layered structure due to the deconfinement which
opens up the interactions between the three-dimensional subspaces at
finite temperature.
\end{abstract}

\newpage
\section{Introduction}
The original idea of Fu and Nielsen in the mid-eighties of the last
century consisted in proposing  a new way of dimensional reduction
through higher dimensional lattice models with anisotropic couplings
\cite{FN}. Since then in a series of papers the phase diagram of the
higher-dimensional lattice models was studied by mean field and
Monte carlo methods. Besides that a mechanism of producing the
anisotropic couplings was proposed invoking a Randall-Sundrum
space-time metric to the (continuum) higher dimensional
models\cite{RS,DFKK}. The lattice model with anisotropic couplings
which came up could help in understanding the localisation of the
gauge interaction on the brane;  the idea was that the usual
four-dimensional space-time is embedded in a higher dimensional bulk
in which the extra dimensions are subject to confinement. Indeed the
assumed strong coupling dynamics along the extra dimension requires
a non-perturbative study. The numerical study on the lattice has
verified the prediction of a  new phase (layer phase) proposed by Fu
and Nielsen who used mean field methods. In this new phase we
established the existence of a massless photon on the
four-dimensional subspace while at the same time  the extra
dimension is confined \cite{FV}. This confinement is responsible for
the fact that the higher-dimensional space is layered-like and the
interactions are confined in the four-dimensional space-time slices.

It is worthwhile to mention that in the Dvali-Shifman model a
similar way of thinking has been used in order to achieve  a
localization mechanism on a brane: the assumption of confinement (of
a non-abelian nature)  along one of the dimensions which limits the
dynamics of the model in a  subspace with one dimension less
\cite{DvS}. It is possible to generate gravitationally this
localization mechanism for the trapping  of the charged fields,
under a non-abelian gauge field, on a 3d submanifold (brane) using
the non-minimal coupling of gravity with a scalar field \cite{FP}.
The result is a spontaneously broken phase on the brane (higgs
phase) and a confining (symmetric phase) in the transverse
directions (bulk).

In this work we consider the  following exercise: we assume a
five-dimensional U(1) lattice model with anisotropic couplings at
finite temperature. Our intention is to discover the fate of the
layer phase as we switch on the temperature to non zero values. By
means of numerical methods we study the phase diagram and our main
result is that for $T \ne 0$ the layer phase becomes a deconfined
phase with new properties. In other words, the confining extra
dimension that is detected at $T=0$, becomes deconfined for non zero
values of the temperature and the system is lacking the
four-dimensional layered structure. \footnote{Our results contradict
the prediction based on a Variational Cumulant Expansion by the
authors of ref.~ \cite{HCFu} for which the layer phase of the
anisotropic lattice gauge models persists at high temperature.}

The previous result is suggestive for the behavior of a
multidimensional anisotropic gauge-higgs model at finite temperature
\footnote{For the analysis of anisotropic gauge-higgs models at
zero temperature see the refs.~\cite{DFKKN,AnisoH}.}.
 It is well known that at high temperature the higgs phase turns into a
symmetric phase. We expect then that the layer higgs phase
disappears at this temperature. The new phase is probably a
multidimensional high temperature symmetric phase:  a hot
multidimensional world in the "quark-gluon plasma" phase. In this
rather  hypothetical scenario the Universe starts as a hot
multidimensional system that cools down,  passes through a series of
phase transitions and ends up to a brane Universe at  zero
temperature.

To explore the phase diagram of the anisotropic 5D U(1) gauge model
we have to understand first the phase structure of the 4D U(1) gauge
model at high temperature. The study of the phase structure of
lattice electrodynamics in three and four dimensions at zero
temperature, using the topological excitations of the theory
(monopoles) was first performed by the authors of ref.~\cite{DeGT}.
Computer simulations for the 3D compact QED at finite temperature
were performed in ref.~\cite{CIS} for the deconfinement transition
from the monopole anti-monopole point of view. In three dimensions
and zero temperature the theory is confining for all values of the
coupling constant and a monopole and anti-monopole plasma is
responsible for the permanent confinement of oppositely charged
particles. For non-zero temperature the binding of monopoles and the
formation of magnetic dipoles lead  to loss of confinement. The
dipole plasma can not sufficiently screen the field created by
electric currents and the screening mass vanishes.

In four dimensions the point-like topological excitations become
one-dimensional objects (strings of monopole currents). Again, in
the zero-temperature case and for small values of the coupling
constant $\beta$ there is a large number of monopole loops winding
around the system and causing disorder. As a consequence, if an
external field is applied it will be shielded after a small
penetration. For large values of $\beta$, on the other hand, the
situation is different. A long distance penetration of the external
field is observed, accompanied by the renormalization of the
magnetic charges due to the monopole currents. The case of 4D
compact U(1) gauge theory at finite temperature was studied
separately in \cite{Berg} and \cite{FdV}. The authors of reference
\cite{Berg} reported a second order phase transition to a Coulomb
phase for $L_{t}\geq 4$, with critical exponents consistent with 3d
Gaussian values and no obvious dependence on $L_{t}$. A different
picture emerged in ref.~\cite{FdV}, where among other things the
disappearance of the Coulomb phase for all values of
$T(\equiv\frac{1}{L_{t}})\neq0$ was predicted. Instead of a Coulomb
phase we are left with a spatial confining - temporal Coulomb phase
for all temperatures.

 In this paper, we are not going to present a detailed study of the nature of the
phase transitions; however, some of our findings seem to indicate
the absence of a Coulomb phase for all temperatures different from
zero for the case of four dimensional compact QED (as long as the
condition $L_{t}<<L_{s}$ is satisfied). We go one step further and
examine a finite temperature scenario in five dimensions through the
anisotropic U(1) gauge model with couplings $\beta$ and $\beta^{'}$.
The connection of this model with the brane model scenarios makes it
an ideal candidate for  the study of the brane models in the
non-zero temperature case. We are mostly interested to discover if
some of the most promising characteristics of this model survive in
the high-temperature regime. In what follows we will try  to give a
brief  summary of our findings through the description of the
limiting cases of our model.

For $\beta^{'}=0$ we  obtain the  four dimensional QED at finite
temperature. From the study  of the system with volume
$V_{4D}=L_{t}\times L_{s}^{3}$ and $L_{t}=2,4$ and 6 we come to the
conclusion that, although phase transitions seem to appear for
finite $L_s$, they are actually  finite-size effects and disappear
in the limit  $L_{s}\rightarrow\infty$. For all values of $L_{t}\neq
0$ the Coulomb phase gives its place to a temporal Coulomb - spatial
confining phase (deconfining phase) \cite{Borgs}. The  Coulomb phase
is recovered only at $T=0$.

For $\beta^{'}\neq 0$ and $L_{t}=2$ we have a five dimensional,
anisotropic model in  a high temperature state. The zero temperature
model  has been already  studied and it is characterized by three
distinct phases \cite{DFKK,FV,DFV,KAN}. A five dimensional confining
phase, a 5D Coulomb phase and the layer phase where the system is
confining along the fifth direction  while, along the four remaining
directions, it exhibits the  Coulomb behaviour with a massless
photon. These characteristics change when the temperature becomes
non-zero. We observed the replacement of the layer phase by a
deconfining phase, due to the same mechanism responsible for the
disappearance of the Coulomb phase in four dimensions. The behavior
of the system in the time direction is coulombic and confining in
the remaining four directions.

Our work is organized as follows. In section 2 we present the action
of the model and the observables, the helicity modulus and the
Polyakov line, that we use in order  to characterize the phase
diagram of the model. In section 3 we analyze the system in the
three limiting cases: 5D anisotropic for $T=0$, $\beta^{'}=0$ and
$L_{t}=1$. Finally in section 4 we present the phase diagram for the
5D anisotropic model at finite temperature and in particular for
$L_{t}=2$.  \footnote{We recall that $T\equiv 1/L_{t}$ in lattice
units.}

\section{The model}
\subsection{Definition}
The five dimensional anisotropic U(1) gauge model with two couplings,
$\beta$ and $\beta^{'}$, at finite temperature is defined as :

\begin{displaymath}
 S^{5D}_{gauge}=\beta\sum_{x,1 \leq \mu < \nu \leq 3}(1- Re(U_{\mu\nu}(x))+
\beta\sum_{x,1 \leq \mu \leq 3}(1- Re(U_{\mu t}(x)) +
\end{displaymath}
\begin{equation}
\beta^{'}\sum_{x,1 \leq \mu \leq 3}(1- Re(U_{\mu 5}(x))+
\beta^{'}\sum_{x}(1 - Re(U_{t5}(x))
\end{equation}
where
\begin{displaymath}
 \begin{array}{ccc}
U_{\mu\nu}(x)& = & U_{\mu}(x)U_{\nu}(x+a_{s}\hat{\mu})U_{\mu}^{\dagger}(x+a_{s}\hat{\nu})U_{\nu}^{\dagger}(x)\\
U_{\mu t}(x) & = & U_{\mu}(x)U_{t}(x+a_{s}\hat{\mu})U_{\mu}^{\dagger}(x+a_{t}\hat{t})U_{t}^{\dagger}(x)\\
U_{\mu 5}(x) & = & U_{\mu}(x)U_{5}(x+a_{s}\hat{\mu})U_{\mu}^{\dagger}(x+a_{5}\hat{5})U_{5}^{\dagger}(x)\\
U_{t 5}(x) & = &
U_{t}(x)U_{5}(x+a_{t}\hat{t})U_{t}^{\dagger}(x+a_{5}\hat{5})U_{5}^{\dagger}(x)
\end{array}
\end{displaymath}
are the plaquette variables defined on the 4d-subspaces \{($\mu, \nu
= 1,2,3$) - t\} and planes containing an extra, fifth dimension
($x_{5}$).  With an obvious noatation we call these
plaquettes as $P_{s},P_{st},P^{'}_{s5}$ and $P^{'}_{t5}$.\\
The link variables are defined as
\begin{displaymath}
 U_{\mu}(x)=e^{i\theta_{\mu}(x)},\quad U_{t}(x)=e^{i\theta_{t}(x)}\mbox{
and}\quad U_{5}(x)=e^{i\theta_{5}(x)}
\end{displaymath}

  Let us make the notation clear. The
action  is defined in an Euclidean lattice volume, namely
 $V=L_{t}\times L_{s}^{3}\times L_{5}$ in lattice units. $L_{t}$ is the compactified temporal
dimension which is related to the temperature through the
relationship
\begin{equation}
 T=\frac{1}{L_{t}a_{t}}
\end{equation}
We denote with $a_t$ the lattice spacing,  $L_{t}$ is an integer number,
$L_{s=1,2,3}$ are the usual
 ''infinite'' space dimensions and finally $L_{5}$ is  an extra, fifth dimension, which we
consider to be infinite and  equal to $L_{s}$. We assume periodic
bourdary conditions for the U(1) gauge field in all directions. The
proclaimed anisotropy of the model has nothing to do with the
''time'' direction. In  this model the lattice spacings
$a_{s}$,$a_{t}$ are equal. The  anisotropy is introduced through the
interaction along the extra direction. So, we have
\begin{math}
 a_{s}=a_{t}\equiv a \mbox{ and } a_{5}\neq a
 \end{math}
where $a_{5}$ is the lattice spacing related to the extra dimension.

 In our model  the gauge couplings $\beta$ and $\beta^{'}$ are generally independent from each other and
the coordinates. The lattice spacing is determined
from the value of the couplings $\beta$ and $\beta^{'}$. In some
cases we can have a coordinate dependence and it is possible to
relate them with extra fields, as in the brane
model~\cite{RS,DFKK,KT}. In terms of the continuum fields the link
angles, $\theta_{M}$, can be written as :
\begin{displaymath}
 \theta_{M}(x)=a_{M}\bar{A}_{M}(x)
\end{displaymath}
where  $\bar{A}_{M}(x)$ are the gauge potentials~\cite{DFKK,DFKKN}
and with $M$ we denote $M=(t,\mu,5)$. In the na\"{\i}ve continuum
limit
\begin{math}
 (a,a_{5}\rightarrow 0)
\end{math}
we define:
\begin{displaymath}
 \beta=\frac{a_5}{g_{5}^{2}} \mbox{ ~~and~~ } \beta'=\frac{a^2}{g_{5}^{2}a_{5}}
\end{displaymath}
where $g_{5}$ is the bare five-dimensional coupling constant for the
gauge field. The resulting continuum action takes the standard form:
\begin{displaymath}
 S_{gauge}=\int d^{5}x\frac{1}{g_{5}^{2}}\overline{F}_{MN}^{2}
 \mbox{ , }\quad
 \overline{F}_{MN}=\partial_{M}\overline{A}_{N}-\partial_{N}\overline{A}_{M}.
\end{displaymath}
Note that $g_{5}^{2}$ has dimensions of length and is related to a
characteristic scale for five dimensions. The previous expression
does not exhibit any anisotropy at all. However, the results that we
present below indicate that the anisotropy may survive in the
continuum limit.

\subsection{Observables}
We now proceed to the introduction of the observables,  i.e. the
gauge invariant quantities which are used for the study of the
model.
\subsubsection{The helicity modulus}
Among  the quantities used to distinguish the various phases and the
respective phase transitions in a statistical model the ones that
attract the most attention are the so called order parameters. Their
great significance comes from the fact that they display completely
different behavior between the various phases. Their ''thermal
average'' is zero on the one side of the transition and moves away
from zero on the other side. For the case of a confining-Coulomb
transition a quantity with the properties of an order parameter is
the helicity modulus ($\bf{h.m}$). It was first introduced in the
context of lattice gauge theories by P.de Forcrand and M. Vettorazzo
and it characterizes the responce of a system to an external flux.
It is zero in a confining phase and nonzero in a coulombic one\cite{FdV}.\\
Let us  consider our five dimensional system with
\begin{math}(L_{\mu},L_{\nu},L_{\rho},L_{t},L_{5})\end{math} and
let us choose a particular orientation, for example, $(\mu,\nu)$.
Through the remaining orthogonal directions it is defined  a stack
of $L_{\rho}\times L_{t}\times L_{5}$ plaquettes parallel to the
$(\mu,\nu)$ orientation. In order to  study the response of the
system to an external static field we assume the presence of an
external flux $\Phi$ through this stack of plaquettes. By a suitable
choice of variable transformations we can spread the flux
homogeneously over the parallel planes. In other words, we can add
the constant value of
\begin{math}\Phi_{P}=\frac{\Phi}{L_{\mu}L_{\nu}}\end{math}
 to each of the plaquettes of the
given ($\mu,\nu$) orientation.
 Also we can impose an external flux by changing the boundary links using
twisted boundary conditions instead of using  pure
periodic~\cite{FdV,Cardy}. The partition function, in the presence
of the external flux, is:
\begin{equation}
Z(\Phi)= \int D\theta e^{-S(\theta;\Phi)}
\end{equation}

\begin{displaymath}
 S(\theta;\Phi)= -\beta\sum_{(\mu \nu)planes} \cos(\theta_{\mu\nu}+\frac{\Phi}{L_{\mu}L_{\nu}})
 -\beta\sum_{(\overline{\mu\nu})planes}\cos(\theta_{\overline{\mu\nu}})
\end{displaymath}
\begin{displaymath}
-\beta\sum_{x,1 \leq \mu \leq 3}\cos(\theta_{\mu t}(x))-\beta^{'}\sum_{x,1 \leq \mu \leq 3}\cos(\theta_{\mu 5}(x))
\end{displaymath}
\begin{equation}
 -\beta^{'}\sum_{x}\cos(\theta_{t5}(x))
\end{equation}
where \begin{math}\sum_{(\mu \nu)planes}\end{math} is the sum over
the plaquettes of the given orientation $(\mu,\nu)$, containing the
flux and \begin{math} \sum_{(\overline{\mu\nu})planes}\end{math} its
complement, consisting of all the plaquettes that remained unchanged
(plaquettes belonging
to the other planes).\\
>From  the partition function we can obtain the flux dependent free
energy
\begin{equation}
 F(\Phi)=-\ln(Z(\Phi))=-\ln\left(\int D\theta {e^{-S(\theta;\Phi)}}\right)
\end{equation}
An important observation is that the partition function $Z(\Phi)$ of equation (3) and hence
the flux free energy is clearly $2\pi$ periodic . So, the extra flux we impose on the system
is defined only $\mbox{mod}(2\pi)$.\\
In the confining phase the flux free energy $F(\Phi)$ is constant in
the thermodynamic limit because the correlation length and the
effect of the external flux through the twisted boundary links is
exponentially decreasing. On the contrary, in the Coulomb phase we
have an infinite correlation length,  so the influence of the
twisted boundary conditions is extended to the full extent of the
system.  As a result we have a nontrivial dependence of $F(\Phi)$ by
the
external flux $\Phi$. \\
The helicity modulus is defined as
\begin{equation}
 h(\beta)= \left. \frac{\partial^{2}F(\Phi)}{\partial \Phi^{2}}\right| _{\Phi=0}
\end{equation}
 and it gives  a measure of the curvature of the free energy profile around
$\Phi=0$. From the above equation and for various choices  with
respect to  the orientation, due to the anisotropy of the model, we
have:
\begin{equation}
h_{s}(\beta)=\frac{1}{(L_{\mu}L_{\nu})^{2}}\left(\left<\sum_{P_{s}}(\beta\cos(\theta_{\mu\nu}))
\right>-\left<(\sum_{P_{s}}(\beta\sin(\theta_{\mu\nu})))^{2}\right>\right)
\end{equation}

\begin{equation}
 h_{t}(\beta)=\frac{1}{(L_{\mu}L_{t})^{2}}\left(\left<\sum_{P_{st}}(\beta\cos(\theta_{\mu t}))
\right>-\left<(\sum_{P_{st}}(\beta\sin(\theta_{\mu
t})))^{2}\right>\right)
\end{equation}

\begin{equation}
 h_{s5}(\beta^{'})=\frac{1}{(L_{\mu}L_{5})^{2}}\left(\left<\sum_{P_{s5}^{'}}(\beta^{'}\cos(\theta_{
\mu 5})) \right>-\left<(\sum_{P_{s5}^{'}}(\beta^{'}\sin(\theta_{\mu
5})))^{2}\right>\right)
\end{equation}

\begin{equation}
 h_{t5}(\beta^{'})=\frac{1}{(L_{t}L_{5})^{2}}\left(\left<\sum_{P_{t5}^{'}}(\beta^{'}\cos(\theta_{
t 5})) \right>-\left<(\sum_{P_{t5}^{'}}(\beta^{'}\sin(\theta_{t
5})))^{2}\right>\right)
\end{equation}
%\newpage
Now, consider for the moment the classical limit
($\beta\rightarrow\infty$) for the action  (4) where all the
fluctuations are suppressed. In this limit the flux is distributed
equally over all the plaquettes of each plane and it does not change
as we cross the  parallel planes. If we expand the classical action
in powers of the flux, since in the thermodynamic limit the quantity
  $\frac{\Phi}{L_{\mu}L_{\nu}}$ is always small , we find:
\begin{displaymath}
 S_{\mbox{\tiny classical}}(\Phi)=\frac{1}{2}\beta \Phi^{2}\frac{V_{5D}}{(L_{\mu}L_{\nu})^{2}}\mbox{\normalsize + constant}\Longrightarrow
F_{\mbox{\tiny classical}}(\Phi)-F_{\mbox{\tiny classical}}(0)=\frac{1}{2}\beta \Phi^{2}\frac{V_{5D}}{(L_{\mu}L_{\nu})^{2}}
\end{displaymath}
where  $V_{5D}=L_{\mu}L_{\nu}L_{\rho}L_{t}L_{5}$ is the 5D lattice
volume.

The above expression for the free energy, F, holds
 all the way up to the phase transition, where fluctuations are present, if one only replaces
the bare coupling by a renormalized coupling,  $\beta \rightarrow
\beta_{R}(\beta)$ (for details see \cite{Cardy,FdV}):

\begin{equation}
F_{[\mbox{\tiny finite $\beta$}]}(\Phi)-F_{[\mbox{\tiny finite $\beta$}]}(0)=
\frac{\beta_{R}}{2}\Phi^{2}\left(\frac{L_{\rho}L_{t}L_{5}}{L_{\mu}L_{\nu}}\right)
\end{equation}
>From the Eqs. (6) and (11) we have for the ``spatial'' h.m
\begin{equation}
 h_{s}(\beta)\sim \beta_{R}L_{t}
\end{equation}
and following the same steps, we can get the scaling relations for
the remaining quantities:
\begin{equation}
h_{t}(\beta)\sim\beta_{R}\frac{L_{\mu}^{2}}
{L_{t}}
\end{equation}
\begin{equation}
 h_{s5}(\beta^{'})\sim\beta^{'}_{R}L_{t}
\end{equation}
\begin{equation}
 h_{t5}(\beta^{'})\sim \beta^{'}_{R}\frac{L_{\mu}^{2}}
{L_{t}}
\end{equation}
Although the arguments presented above are based mainly on the classical approach, this is indeed the case
in the Coulomb phase and the helicity moduli applied for the five dimensional system behave exactly as
the above equations predict.

\subsubsection{Polyakov loop (or Wilson line)}
 For the  evaluation of  the potential between a static
quark-antiquark pair at zero temperature, the study of the ground
state expectation value of the Wilson loop for large Euclidean times
is needed. When the temperature is non zero ($L_{t} << L_{s}$  as
opposed to the former case) the same information is obtained by
using  a different quantity which is the Polyakov loop or the Wilson
line. It consists of the product of link variables along
topologically non-trivial loops, winding around the time direction
due to periodic conditions.
\begin{displaymath}
 P_{t}(\vec{n},n_{5})=\prod_{n_{t}=1}^{L_{t}} U_{t}(\vec{n},n_{t},n_{5})
\end{displaymath}
\begin{equation}
P_{t}=\frac{1}{L_{s}^{4}}\sum_{(\vec{n},n_{5})}P_{t}(\vec{n},n_{5})
\end{equation}
where \begin{math}\{n\}\epsilon Z^{5}\end{math} denotes a lattice site.\\
Physically, the expectation value of the Polyakov loop determines the free energy of a
system with a single, static heavy quark, measured relative to the vacuum:
\begin{equation}
\left<|P_{t}|\right>=e^{-L_{t}F_{q}}
\end{equation}
where $|P|$ is the absolute value of P and $<\dots>$ the statistical
average value evaluated using the action of equation (1). The above
relation holds even in the presence of finite mass quarks coupled to
the gauge potential with the only difference that in that case the
expectation value has to be calculated using the finite temperature
action that includes the dynamical fermions.

The Polyakov loop is somewhat the world line of a static quark in a Wilson loop and
that suggests that the free energy of a quark-antiquark  pair located at
$(\vec{n_{1}},n_{5_{1}})$ and $(\vec{n_{2}},n_{5_{2}})$ respectively is given by the
correlation function of two such loops, with bases at the aforementioned points and having
opposite orientations. Consequently we have:
\begin{equation}
\left<P_{t_{1}}(\vec{n_{1}},n_{5_{1}})P_{t_{2}}^{\dagger}(\vec{n_{2}},n_{5_{2}})\right>=
e^{-L_{t}F_{q\bar{q}}(\{\vec{n_{1}},n_{5_{1}}\};\{\vec{n_{2}},n_{5_{2}}\})}
\end{equation}
keeping $L_{t}$ constant.\\
For large distance separation of the quark-antiquark pair and assuming that the correlation
functions satisfy clustering, the above expression reduces to
\begin{equation}
\left<P_{t_{1}}(\vec{n_{1}},n_{5_{1}})P_{t_{2}}^{\dagger}(\vec{n_{2}},n_{5_{2}})\right>
 \rightarrow |\left<P_{t}\right>|^{2}
\mbox{ for } |\hat{R}|\rightarrow \infty\quad (\mbox{where }
\hat{R}=\{\vec{n_{1}},n_{5_{1}}\}-\{\vec{n_{2}},n_{5_{2}}\})
\end{equation}
which is just the self-energy of two isolated quarks.\\
In the confinement phase the correlation function of the Polyakov
loop decays exponentially for large distances:
\begin{equation}
\left<P_{t}(0)P_{t}^{\dagger}(\hat{R})\right>\sim e^{-L_{t}\sigma|\hat{R}|}
\end{equation}
giving a linear potential with string tension $\sigma$ and $F_{q\bar{q}}\simeq\sigma|\hat{R}|$.\\
The flux free energy $F_{q\bar{q}}$ increases, in general, for large
separation of the quarks in the confining phase, giving eventually
$<|P_{t}|>=0$ and $F_{q\bar{q}}=\infty$ in the thermodynamic limit.
We interpret $<|P_{t}|>=0$ as a signal for confinement. If we have
$<|P_{t}|> \neq 0$, then the free energy of the static quark-antiquark
pair tends to a constant value, for large separation of the heavy
charges, as shown in Eqs.~(18) and (19), and this is a signal
for deconfinement. In other words, the expectation value of the
temporal Polyakov loop serves as an order parameter in finite
temperature gauge theories.

\section{Three limitting cases}
\subsection{The zero temperature case}

\begin{figure}[hb!]
\begin{center}
\includegraphics[width=8 cm,angle=270]{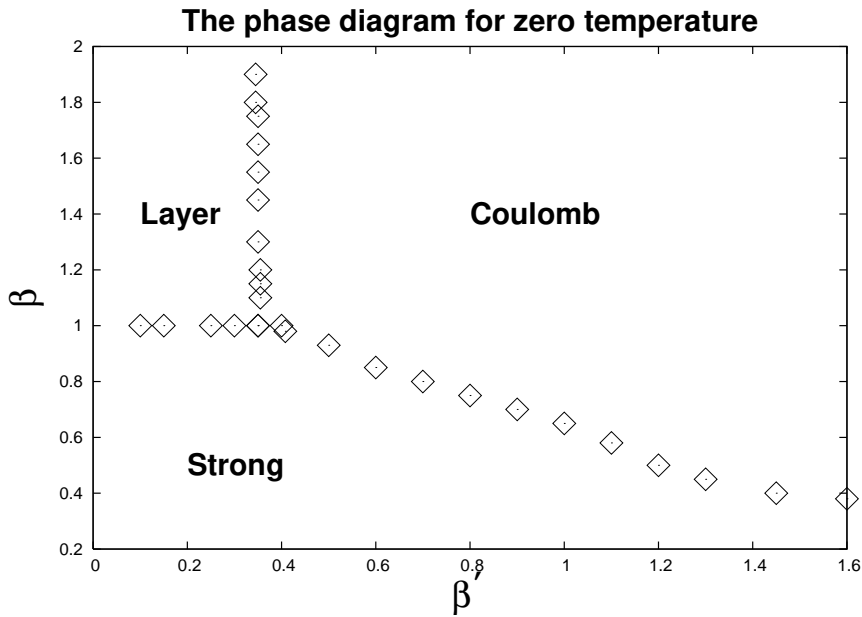}
\end{center}
\caption{The phase diagram for the 5D anisotropic U(1) gauge model
at zero temperature. Three phases are present: Strong confining
phase, 5D Coulomb phase and the Layer phase.} \label{3-1-1}
\end{figure}

The five dimensional anisotropic U(1) gauge model, at zero
temperature, was first introduced by Fu and Nielsen \cite{FN} as  an
attempt to offer an alternative way to achieve dimensional
reduction. Since then many numerical investigations of the model
have been made \cite{DFV,KAN}.  As we have already noted in the
Introduction, the interest in this comes from the fact that the
anisotropy of the model produces a new phase, the so called layer
phase, which can serve as a mechanism for gauge field localization
on a brane. We can induce this anisotropy to the gauge coupling
using, for example, the Randall-Sundrum metric background in five
dimensions. The effect of the warp factor from the RS background or
a general anti-de-Sitter ($AdS_{5}$) background on the U(1) gauge
theory is to provide the gauge theory with a different gauge
coupling in the fifth direction(\cite{DFKK}).

In Fig.~1 we present the phase diagram of the theory. It consists
of three distinct phases. For large values of $\beta$ and
$\beta^{'}$ the model lies in a Coulomb phase ($\bold{C}$) on the
5-D space. Now, if one keeps $\beta$ constant and  bigger than one
and at the same time lets  decrease $\beta^{'}$, one will eventually
come across the new phase, the layer phase ($\bold{L}$), where the
forces in four dimensions will still be Coulomb-like but in the
fifth dimension the  confinement is present. For small values of
$\beta$ and $\beta^{'}$ the force is confining in all five
directions and the corresponding phase is the Strong phase
($\bold{S}$). The properties of the three phases can become more
transparent in terms of two test charges. In the Coulomb phase the
force between two heavy charges is 5D Coulomb-like, and becomes the
exact five dimensional Coulomb law in the diagonal line, defined by
\begin{math}\beta=\beta^{'}\end{math} for which  no anisotropy
appears (for a numerical investigation see \cite{FV}). The
completely opposite picture emerges in the Strong phase. There the
force is confining in all five directions giving infinite energy for
the separation of the test charges in any direction. Now, two test
charges in the layer phase will experience a Coulomb force in the
four dimensional layers, with the coupling given by the
four-dimensional coupling $\beta$;   there are strong indications of
the similarity with the usual 4D Coulomb law (see \cite{FV} for
details), while along the fifth direction the test charges will
experience a strong force as the corresponding coupling $\beta^{'}$
takes small values. This means that charged particles in the layer
phase will mainly run only along a layer since in an attempt to
leave the layer in which they belong they will be driven back by a
linear potential, analogous to the one responsible for the quark
confinement. This is the mechanism on which the gauge field
localization scheme is based.

Now we would like to sketch the three phases in terms of the
helicity modulus. In the zero temperature case ($L_{t}=L_{s}\equiv
L_{5}$) we are left with only two possible choices. Instead of the
Eqs.~(7)-(10)  we have:
\begin{equation}
 h_{S}(\beta)=\frac{1}{(L_{\mu}L_{\nu})^{2}}\left(\left<\sum_{P}(\beta\cos(\theta_{\mu\nu}))
\right>-\left<(\sum_{P}(\beta\sin(\theta_{\mu\nu})))^{2}\right>\right)
\end {equation}
\begin{equation}
h_{5}(\beta^{'})=\frac{1}{(L_{\mu}L_{5})^{2}}\left(\left<\sum_{P^{'}}(\beta^{'}\cos(\theta_{\mu
5})) \right>-\left<(\sum_{P^{'}}(\beta^{'}\sin(\theta_{\mu
5})))^{2}\right>\right)
\end{equation}
The first one, $h_{S}(\beta)$, is used to probe the response of the
system to an external flux in the spatial planes (belonging to a 4d
layer) while  the second one, $h_{5}(\beta^{'})$, is used in a
similar way for the planes containing the extra, transverse
direction. \\
(i) In the Strong phase (keeping $\beta^{'}$ constant) the
space-like helicity modulus vanishes (which is  a clear signal of
confinement);  as we approach and eventually pass the phase boundary
it becomes non-zero in the layer phase with a value that approaches
1 as $\beta$ increases further. On the other hand, the transverse
h.m, $h_{T}(\beta^{'})$, remains zero throughout the transition
since both phases exhibit
confinement in the fifth direction. \\
(ii) For the transition between the 5D Coulomb phase and the layer
phase,  $h_{S}(\beta)$ retains a value close to 1 for all values of
$\beta^{'}$, since the four dimensional layers experience already a
4d Coulomb-like phase, while $h_{T}(\beta^{'})$ vanishes for the
layer phase;  as the system  crosses the critical point and enters
the Coulomb phase it grows towards 1 as $\beta^{'}$ increases
further \cite{FV,DFV}.

%\newpage
\subsection{The $\mathbf{\beta^{'}=0}$ case}
On the axis defined by  $\beta^{'}=0$ we consider  the
four-dimensional U(1) model.  In this section our intention is  to
strengthen the arguments given in references \cite{FdV}. We present
numerical results showing that we have a Coulomb phase only for T=0,
in accordance with Fig.~12 of~\cite{FdV}. Our findings contradict
the ones of ref.~\cite{Berg} that stipulates the existence of a
Coulomb phase for $L_{t}\geq 4$ and $\beta\geq \beta_{c}$. The
numerical results presented below show that we have a spatial
confinement phase when the spatial lattice size $L_{s}$ gets big
enough, compared to the temporal size $L_{t}$
\begin{math}(L_{s}\gtrsim 4 L_{t})\end{math}.

This behavior can be understood, following closely the
ref.~\cite{FdV}, using  simple theoretical arguments. In order to
have $h_{s}(\beta)\sim 0$, one must have at least two monopole loops
(far apart) winding around the time direction with opposite
orientations. A non contractible time-like monopole loop can, in
principle, disorder all the spatial planes in the lattice. The
probability to have one such loop passing through a given lattice
site is, roughly, $e^{-m_{\mbox{\tiny mon}}(\beta)L_{t}}$, where
$m_{\mbox{\tiny mon}}(\beta)$ is the monopole mass. So   the
condition to achieve a probability of order 1 for a  system to
containing one (or two)
wrapping monopole loops, is:\\
\begin{equation} L_{s}^{3}\times e^{-m_{\mbox{\tiny mon}}(\beta)L_{t}} \sim 1 \Longrightarrow
L_{s} \sim e^{\frac{L_{t}}{3} m_{mon}(\beta))} \Longrightarrow L_{s} \sim e^{\frac{L_{t}}{3} c\beta}
\end{equation}
since $m_{\mbox{\tiny mon}}(\beta)$ is of order $\beta$. Now,
starting  from the above equation we can make two statements. First,
it predicts a pseudocritical coupling
\begin{math}
 \beta_{c}\sim \mbox{\normalsize log}(L_{s})
\end{math}
which was verified by our measurements and second, that as we go to
smaller temperatures (bigger $L_{t}$), for sufficiently large
$L_{s}$ Eq.~(23) is satisfied; hence the spatial planes becomes
completely disorder and the Coulomb phase disappears.

\begin{figure}[h!]
\begin{center}
\includegraphics[width=7 cm,angle=270]{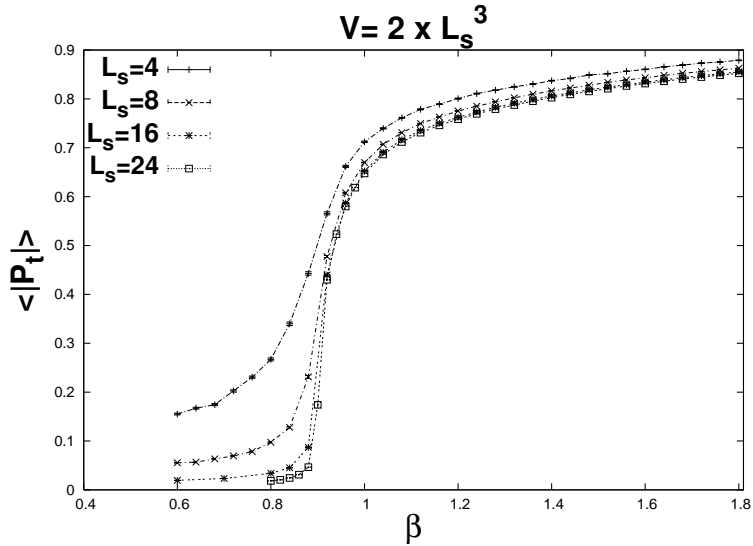}
\end{center}
\caption{The mean value $<|P_{t}|>$ of the temporal Polyakov loop
for $\beta^{'}=0$ and temporal size $L_{t}=2$.
 $<|P_{t}|>$ goes to zero in the confining phase for $L_{s}\rightarrow\infty$.
In the deconfinement phase ($\beta > \beta_{c}(L_{t})$) $<|P_{t}|>$
is non-zero, approaching the value of  one as $\beta$ increases}
\label{3-2-5}
\end{figure}

In Fig.~2 we show the mean value of the temporal Polyakov loop
$P_{t}(\beta)$ for $L_{t}=2$. There is an obvious continuous phase
transition from the confining phase
\begin{math}(\beta \leq \beta_{c}\simeq 0.90)\end{math}, where $<|P_{t}|>$ is zero, to the
deconfining phase where $<|P_{t}|>$ approaches  the value of one. In
the confining phase  the free energy of a single static charge,
relative to the vacuum, goes to infinity with $L_{s}$ while   it
gets  a positive value in the deconfining phase which vanishes as
 $\beta$ increases. The mean value of the temporal Polyakov loop
remains always non zero in the finite temperature phase as we switch
on  $\beta^{'}$. This is the case presented in   section 4. However
 this order parameter does not help us to characterize further
the nature of the different phases. We  arrived at the conclusion
that the helicity modulus is a more promising quantity to study in
detail the phase diagram.

\begin{figure}[!hb]
\begin{center}
\subfigure[\quad]
{\includegraphics[scale=0.30,angle=270]{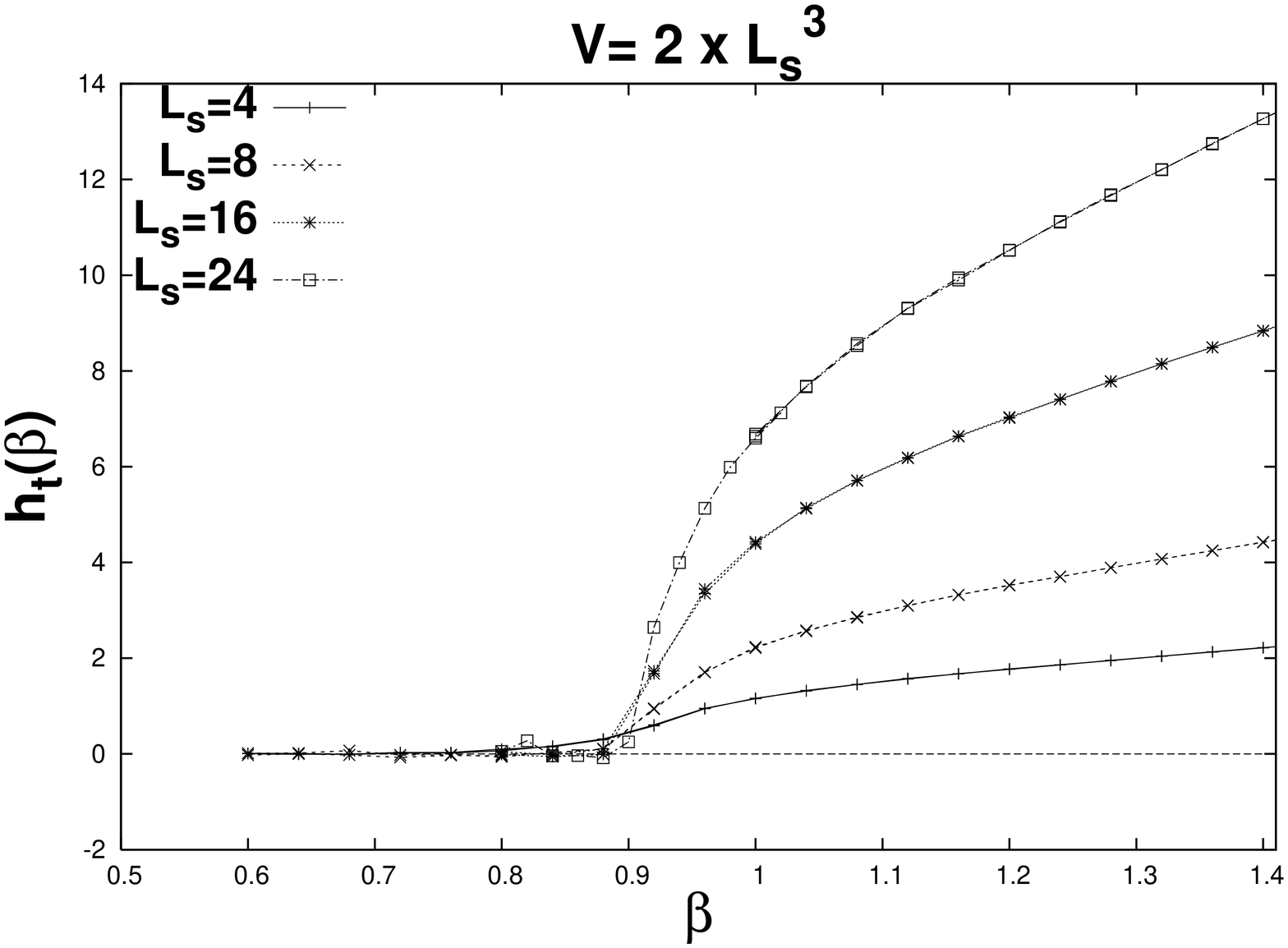}}\quad
\subfigure[\quad]
{\includegraphics[scale=0.30,angle=270]{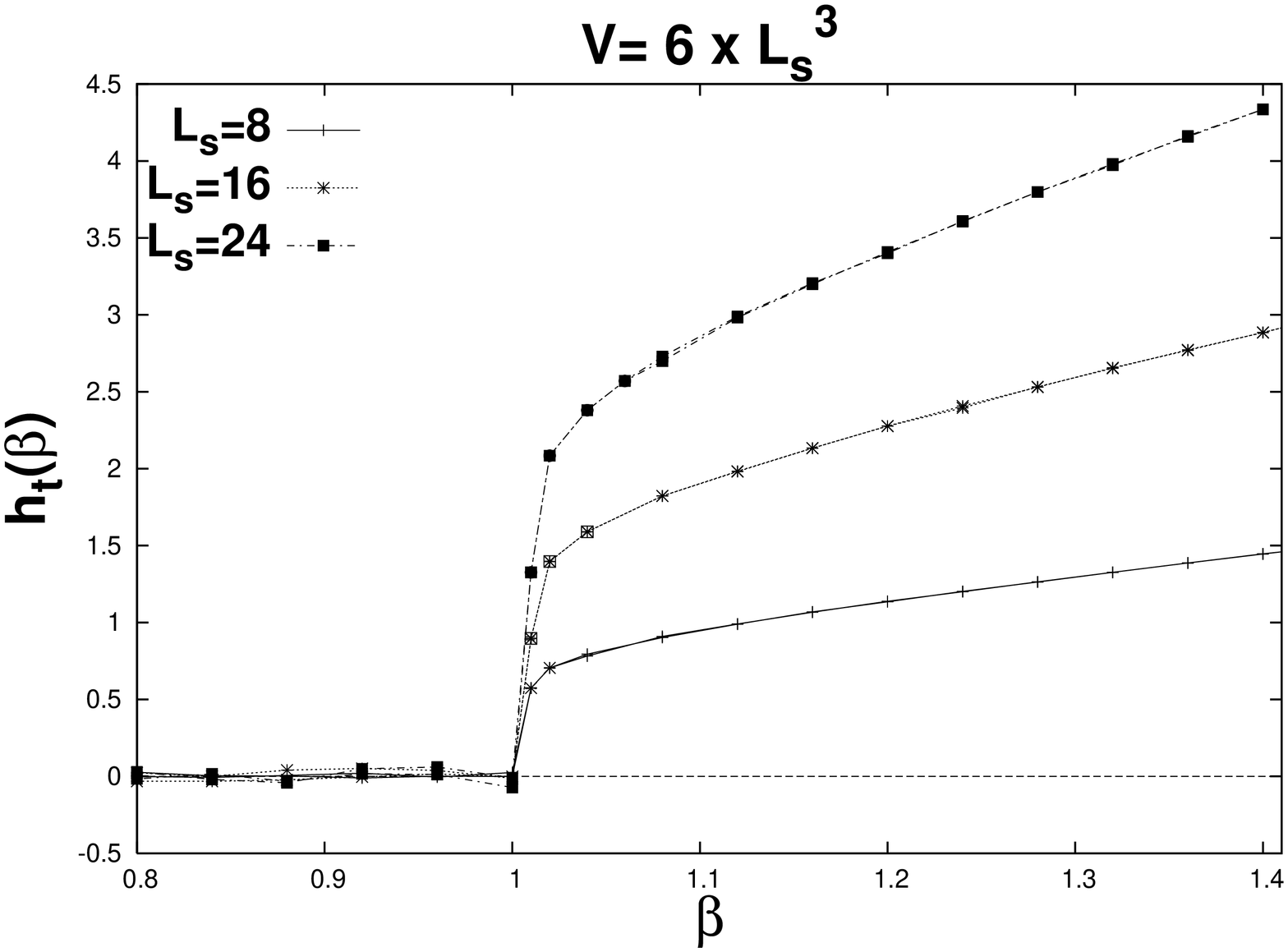}} \caption{
The temporal helicity modulus $h_{t}$ for $L_{t}=2$ (a) and
$L_{t}=6$ (b) and $\beta^{'}=0$. Results from three different
volumes are present. The value of $h_{t}$ increases with $L_{s}$ in
the deconfine region, for $\beta$ bigger than a critical value,
again with accordance with the scaling predictions of section
(2.2.1). The transition for $L_{t}=2$ is continuous, as opposed to
the $L_{t}=6$ case, where we have a discontinuous behavior. }
\end{center}
\end{figure}
In figure 3 we present our results for the temporal helicity modulus
$h_{t}(\beta)$ for two different ``temperatures'' $L_{t}=2$ and
$L_{t}=6$. The temporal h.m is zero in the confining phase for
$\beta < \beta_{c}(L_{t})$ and non-zero for $\beta \geq
\beta_{c}(L_{t})$ in the deconfining phase, indicating coulombic
behavior. The signal for $h_{t}(\beta)$ in the deconfining phase
($\beta \geq \beta_{c}$) is being enhanced with increasing $L_{s}$,
following the scaling relation
\begin{math}h_{t}\sim \frac{L_{s}}{L_{t}}\end{math}.
The transition point has only a weak dependence from the lattice
volume showing convergence to a critical value $\beta_{c}(L_{t})$
with $L_s$. We see that $\beta_{c}(L_{t})$ tends to smaller values
as $L_{t}$ decreases.\footnote{$\beta_{c}(L_{t}=2)=0.9008(3)$,
$\beta_{c}(L_{t}=4)=1.00340(1)$ and
$\beta_{c}(L_{t}=6)=1.0094491(1)$. The results are  taken from
ref.~\cite {FdV}.} Another noticeable difference is the behavior of
$h_{t}$ in the critical region. For $L_{t}=2$ , $h_{t}(\beta)$ goes
continuously to zero when $\beta$ approaches $\beta_{c}$ from above.
For $L_{t}=6$, on the other hand,  the $h_{t}$ has an obvious
discontinuity as $\beta$ approaches $\beta_{c}$ and the volume
increases. This behavior indicates a different order for the phase
transition, a second order phase transition for $L_{t}=2$ and a
first order for $L_{t}=6$ (for details see ref.~\cite {FdV}).

In Fig.~4 we show the spatial helicity modulus $h_{s}(\beta)$ for
$L_{t}=2$ and spatial lattice sizes $L_{s}=4,8,16$ and 24. The
spatial helicity modulus is zero for $\beta$ smaller than a critical
value $\beta_{c}(L_{s})$ that depends strongly on $L_{s}$. We shall
refer from now on to $\beta_{c}(L_{s})$ as the pseudocritical value,
to distinguish it from the real critical value of  $\beta$ that
comes from the temporal helicity modulus $h_{t}(\beta)$. For
\begin{math}\beta\geq \beta_{c}(L_{s})\end{math}, $h_{s}$ takes non-zero values, increasing
linearly as $\beta$ takes bigger values . On the other side, the
magnitude of this quantity decreases according to the ratio $\sim
\frac{L_{t}}{L_{s}}$, as we increase $L_{s}$ and tends to zero for
$L_{s}\rightarrow \infty$. The pseudocritical value
$\beta_{c}(L_{s})$ increases with $L_{s}$ as $\log(L_{s})$
\cite{FdV} and the ratio $\frac{\beta_{c}(L_{s})}{\log(L_{s})}$
tends to the value 0.5 for $L_{s}\geq 16$. In this way
$\beta_{c}(L_{s})$ goes to infinity when $L_{s}\rightarrow \infty$.
As a result the spatial h.m is always zero in the infinite volume
limit and as a consequence we have a spatial confining phase.
\begin{figure}[ht!]
\begin{center}
\includegraphics[width=7 cm,angle=270]{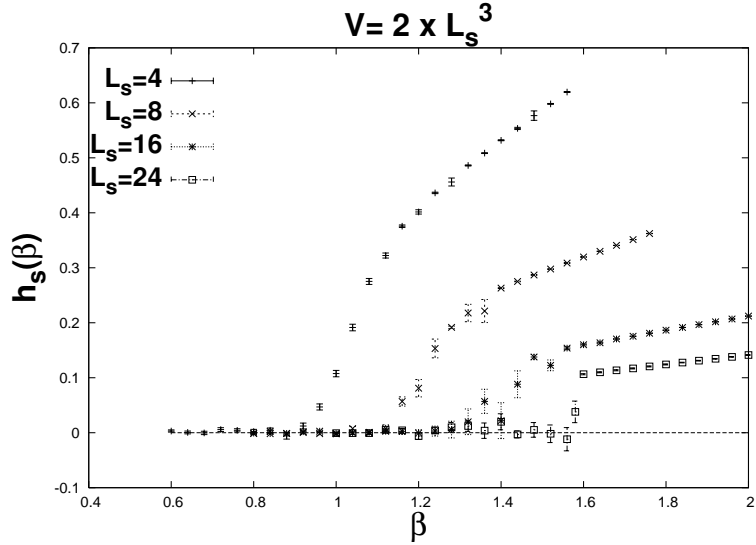}
\end{center}
\caption{The spatial helicity modulus $h_{s}$ for $L_{t}=2$ and
$\beta^{'}=0$, versus the four dimensional coupling $\beta$. The
pseudo critical value of $\beta$ increases very fast, towards an
infinite value, as the spatial lattice size $L_{s}$ increases. The
shift of the transition region to bigger values of $\beta$ is
obvious even in  the smaller volumes.} \label{3-2-1}
\end{figure}
%\newpage

Finally we study the spatial helicity modulus for $L_{t}=4$ and
$L_{t}=6$. The results are shown in Figs.~5(a)($L_{t}=4$) and
5(b)($L_{t}=6$). We have, in general, the same situation as for
$L_{t}=2$. There is a pseudo-critical value of $\beta$ that moves to
larger values as $L_{s}$ increases showing  a strong dependence on
 $L_{t}$. The signal for $L_{t}=4$ is clear only for $L_{s}\geq1
6$. For $L_{t}=6$ it seems that $L_{s}=16$ is not enough but for
$L_{s}\geq 24$ we get  a clear displacement of $\beta_{c}(L_{s})$ to
the right. In the region $\beta \leq \beta_{c}(L_{s})$ the spatial
helicity modulus is zero (confining region). For $\beta > \beta_{c}$
the spatial h.m scales with $\sim \frac{L_{t}}{L_{s}}$ and tends to
zero for $L_{s}\rightarrow\infty$. If we examine the $L_{t}=8$ case
for example, we would probably need volumes bigger than $8\times
32^{3}$ in order to get a  clear picture of  the behavior of the
system. From the previous observations we can say that
$h_{s}(\beta)$ is zero for every value of $\beta$ in the infinite
volume limit, and consequently, we have spatial confinement for all
temperatures different from zero.

We conclude that the phase diagram  on the $\beta$,
$T=\frac{1}{L_t}$ plane has three phases: a confining phase for
$\beta<\beta_c(L_t)$, a temporal Coulomb - spatial confining phase
for $\beta>\beta_c(L_t)$ \footnote{This phase is usually called
deconfining phase.} and the pure Coulomb phase for
$L_{t}\rightarrow\infty$ and $\beta>\beta_c$ \cite{FdV,Borgs}.
\begin{figure}[!h]
\begin{center}
\subfigure[\quad]
{\includegraphics[scale=0.30,angle=270]{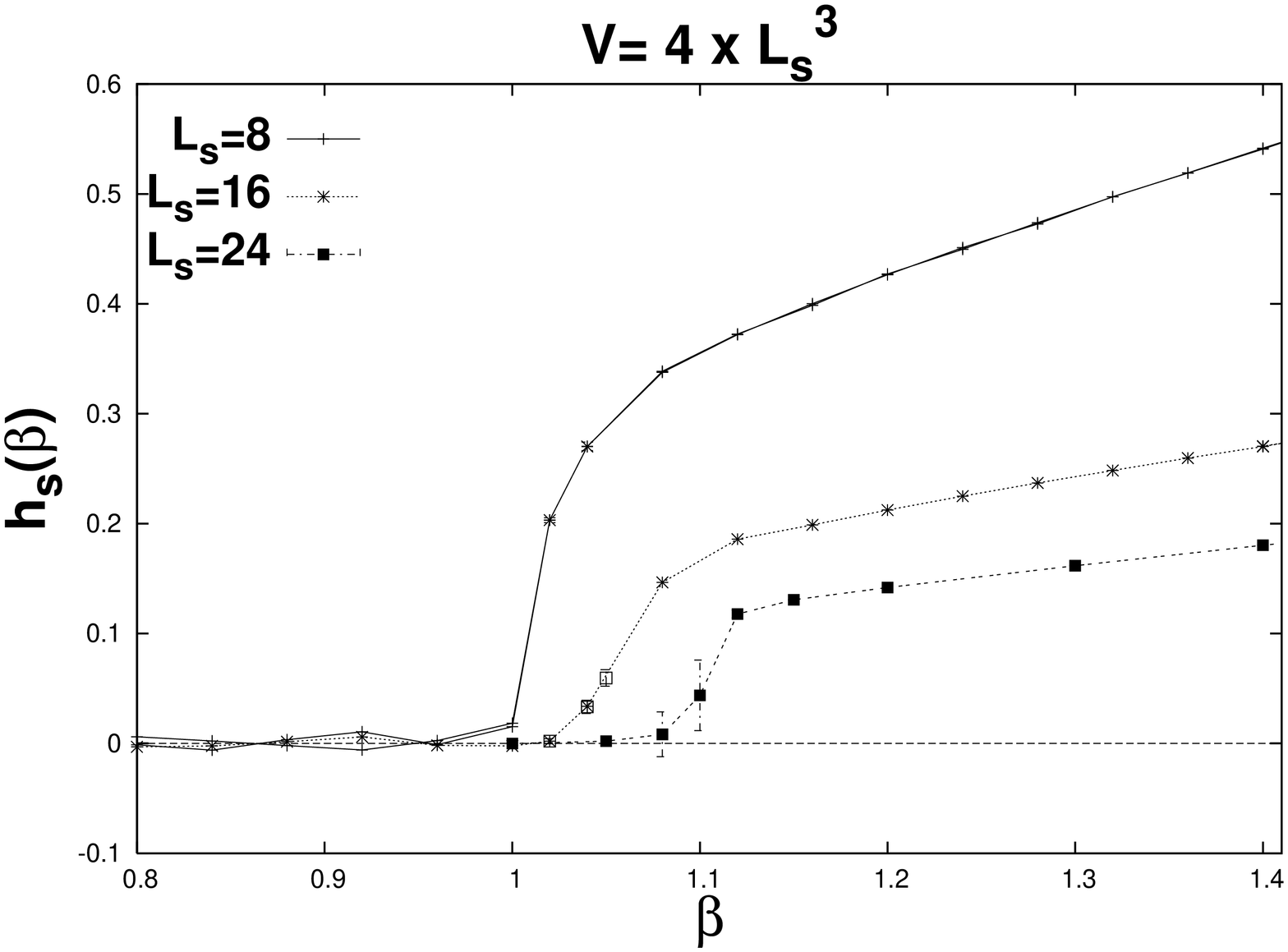}}\quad
\subfigure[\quad]
{\includegraphics[scale=0.30,angle=270]{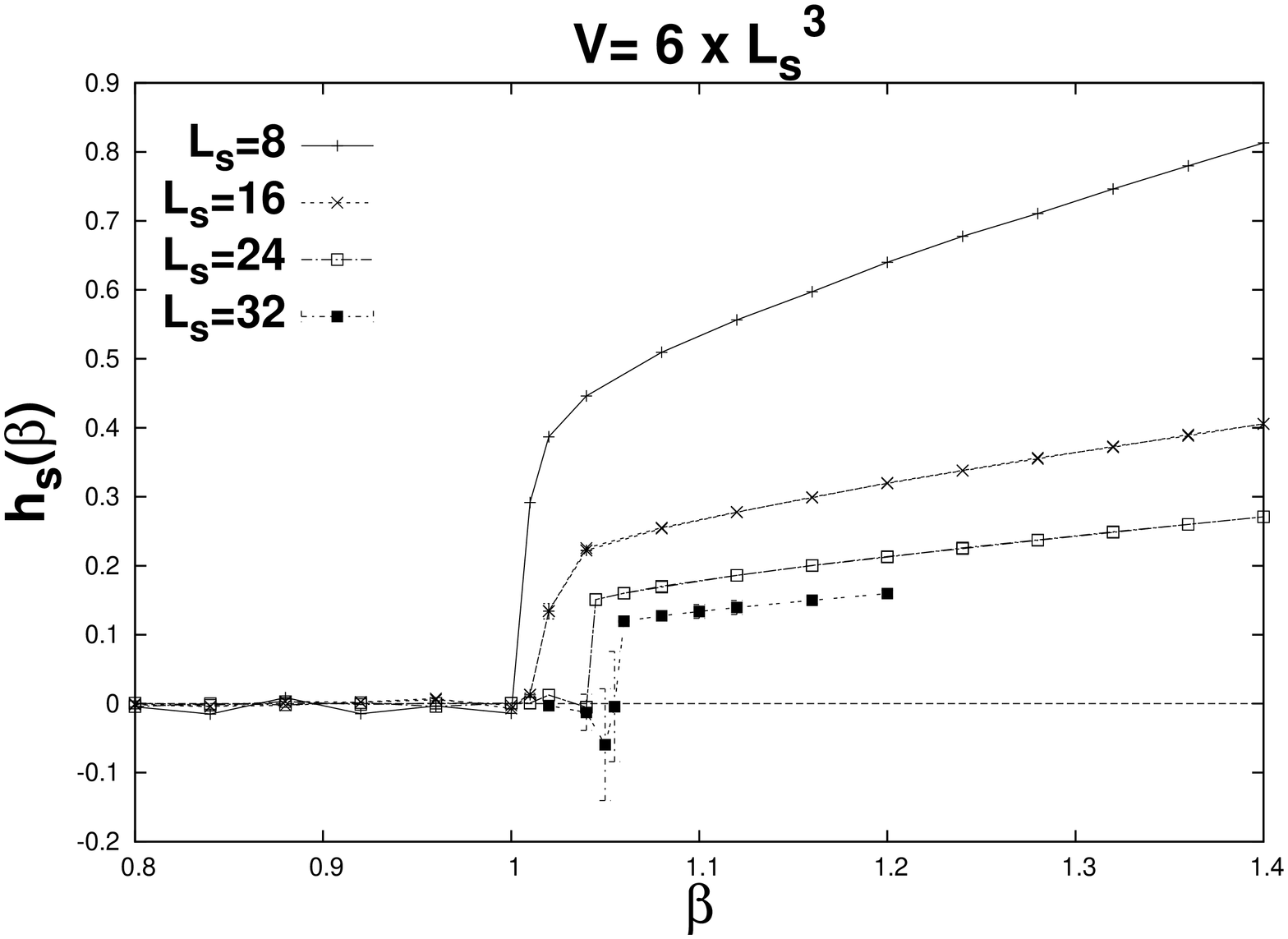}} \caption{In
figures (a) and (b) we present the spatial h.m ($h_{s}$) for
$\beta^{'}=0$ and different temporal sizes, $L_{t}=4$ and $L_{t}=6$,
for a variety of spatial volumes. The size of $h_{s}$ decreases with
$L_{s}$ when  $\beta$ takes values  bigger than the pseudo critical
value, as predicted in section (2.2.1). The transition region moves
clearly to the right as the volume increases, in agreement with the
$L_{t}=2$ behavior.}
\end{center}
\end{figure}

%\newpage
\subsection{$\mathbf{L_{t}=1}$}
 In this case the temporal link is a Polyakov
loop by itself and of course it is a gauge invariant quantity. The
temporal plaquette becomes:
\begin{displaymath}
 \theta_{\mu t}(x)=\theta_{\mu}(x)+\theta_{t}(x+\hat{\mu})-\theta_{\mu}(x)-\theta_{t}(x)
\end{displaymath}
The two spatial links cancel each other, so in the U(1) case we get:
\begin{displaymath}
 \theta_{\mu t}(x)=\theta_{t}(x+\hat{\mu})-\theta_{t}(x)
\end{displaymath}
The contribution to the action is:
\begin{equation}
 S_{t}=-\beta\sum_{x,1 \leq \mu \leq 3}\cos(\theta_{t}(x+\hat{\mu})-\theta_{t}(x))
\end{equation}
The same applies for the ``temporal-transverse'' plaquettes and
following the same steps as above we find that their contribution to
the action is :
\begin{equation}
 S_{t^{'}}=-\beta^{'}\sum_{x}\cos(\theta_{t}(x+\hat{5})-\theta_{t}(x))
\end{equation}
 From the above equations it can be observed that the temporal plaquettes
decouple from the space and transverse ones. Equations (24) and (25)
describe a 4D $\bold{XY}$ model with anisotropic couplings
$(\beta,\beta^{'})$. The three spatial links and the fifth,
transverse link form a separate four dimensional anisotropic U(1)
gauge theory with two couplings, $\beta$ and $\beta^{'}$. As a
result the partition function of the model reduces to:
\begin{equation}
 \mbox{\Large Z}_{\left(\bold{L_{t}=1}\right)} = \mbox{\Large Z}_{\mbox{anisotropic  } \bold{4D-XY}}\times \mbox{\Large Z}_{\mbox{anisotropic } \bold{4D-U(1)}}
\end{equation}
and it  describes two independent lattice field theories.

The anisotropic $4D-\bold{XY}$ model, for $\beta^{'}=0$, reduces to
the three dimensional $\bold{XY}$ model which has a second order
phase transition for $\beta=0.4542$ ~\cite{GH}. The phase transition
line continues to the $(\beta,\beta^{'})$ plane for smaller values
of $\beta$ as $\beta^{'}$ increases and  the critical value of
$\beta$ seems to tend asymptotically to the value of 0.1 as
$\beta^{'}$ goes to infinity. \footnote{For example,  for the  4D
$\bold{XY}$ model the critical value is at $\beta=\beta^{'}=0.29(1)$
(see Fig.~6).}

The 4D gauge model for $\beta^{'}=0$ reduces to a three dimensional
U(1) gauge theory which is always in the confining phase. In the
$(\beta,\beta^{'})$ plane we have a critical line which separates
the strong confining phase from the four dimensional Coulomb phase.
If we move  along the diagonal, for example, where
$\beta=\beta^{'}$, we get  the usual weak first order phase
transition for $\beta=\beta^{'}=1.001113$ \cite{Arnold, DFV}.

The above discussion  can be summarized  in the three dimensional
plot of Fig.~6 . The vertical axis is for the temperature given in
terms of the discrete variable $L_{t}$. The upper plane for
$L_{t}=1$ corresponds to ''infinite'' temperature  while the lower
plane for $L_{t}=L_{5}=L_{s}$ corresponds to the zero temperature
case.
\begin{figure}[ht!]
\begin{center}
\vspace{-1 cm} \hspace{-3 cm}
\includegraphics[width=18 cm, height=12 cm, angle=0]{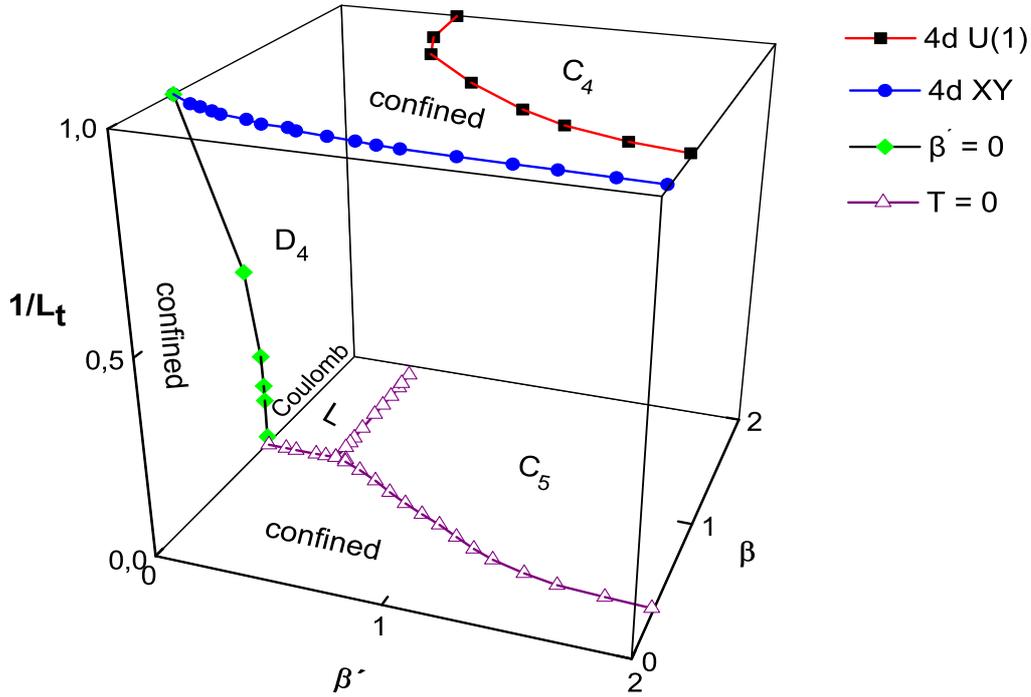}
\end{center}
\vspace{-1.5 cm} \caption{Three dimensional phase diagram of the
model, the vertical axis ($1/L_{t}$) represents the temperature. We
present the critical curves for the three limiting cases. $C_4$ and
$C_5$ are the four dimensional and the five dimensional Coulomb
phases respectively. $L$ stands for the layer phase at zero
temperature. $D_4$ is the temporal Coulomb - spatial confining phase
for $\beta^{'}=0$.} \label{3-3-1}
\end{figure}
%\newpage

\section{Study of the phase diagram for $\mathbf{L_{t}=2}$.}
In five dimensions the phase diagram at zero temperature is given in
Fig.~1. \footnote{Lower plane ($1/L_{t}=0.0$), see Fig. 6.} For
$0\leq\beta^{'}<0.40$ and $\beta\approx1.$ there is a critical
horizontal line in the phase diagram separating the 5D strong
confining phase from the layer phase. For $\beta > 1$ and
$\beta^{'}\simeq0.35$ there is a critical vertical line that
separates the layer from the 5D Coulomb phase. Our intention in this
section is to explore the effects of finite temperature on our
system and the  most important, the feasibility (if any ) of a layer
phase, through the study of the changes in  the aforementioned phase
line boundaries and the phases themselves. To that end we move,
first, on the line $\beta^{'}=0.20$ in order to study the
strong-layer phase transition at finite T;  we know that for
$\beta^{'}=0$ (subsection 3.2) there is phase transition for
$\beta\simeq0.90$. Second, we move along  the line $\beta=1.10$, in
order to study the layer-Coulomb phase transition at finite
temperature.\footnote{We refer to the case of the plane $(\beta,
\beta^{'})$ at $1/L_t=0.5$ in Fig.~6.}
 As we will explain in section 4.2 and  using the
Figs.~4 and 7 in order  to have a clear picture of  the behavior
of the system for  bigger values of $\beta$ we need even  bigger
five dimensional volumes than  those that we can presently achieve.

 Using the results presented in  the two following
sections we can argue that the layer phase disappears for $L_{t}=2$
and becomes a deconfined phase with new properties which will be
described below. We can also generalize the arguments and say that
there is no layer phase in finite temperature for any temperature
different from zero. The existence of the layer phase is based
strongly on the existence of the Coulomb phase for $\beta^{'}=0$.
 However there is no Coulomb phase for $\beta^{'}=0$ at
$T\neq 0$ as it is argued in ref.~\cite{FdV}.  We also  confirm this
result (see subsection 3.2).
\subsection{Moving along  the line $\beta^{'}$=0.20}
We begin the investigation of the 5D anisotropic pure U(1) gauge
model at finite temperature with what used to be called  as a 5D
strong-layer phase transition at zero temperature (Fig.~1). We
utilize the helicity modulus $h_{s}(\beta),h_{t}(\beta)$ in order to
bring out the features of the transition and compare  them with the
T=0 and $\beta^{'}=0$ cases. As it is shown in Fig.~3 the first
deviation from the zero temperature case comes from the fact that
now, the transition line boundary between the two phases, is found
at a lower  value of $\beta=0.90$ in contrast with  the value of
$\beta=1.001113$ for $T=0$  case. Another observation is that the
values obtained here  concerning $h_{t}(\beta)$ are of the same
order of magnitude as the ones, for the $\beta^{'}=0$ case;  the
only difference is the slight movement of the critical region to a
value between 0.85 and 0.90.

Moving now to a discussion of Fig.~7 and the spatial helicity modulus $h_{s}(\beta)$ we
encounter many similarities with the results of subsection 3.2 : \\
i) There is a pseudocritical value $\beta_{c}(L_{s})$ for each
lattice size, with $h_{s}$ equal to zero for
$\beta\leq\beta_{c}(L_{s})$, signal of spatial confinement. For
$\beta > \beta_{c}(L_{s})$ the spatial helicity modulus $h_{s}$
increases with $\beta$, as one would expect from a Coulomb phase.
But the transition point moves to higher and higher values of
$\beta$ as the spatial extent of the lattice ($L_{s}$) grows. What
we see here is only a finite size
effect that ceases to exist in the thermodynamic $L_{s}\rightarrow\infty$ limit.\\
ii) The magnitude of $h_{s}(\beta)$, calculated on a single 4d
layer, decreases with $L_{s}$ for the same value of $\beta$ for
$\beta > \beta_{c}(L_{s})$, following the ratio
$\sim\frac{1}{L_{s}}$. So we expect, as in the 4d case for
$\beta^{'}=0$, that the spatial helicity modulus tends to zero for
all values of $\beta$ as $L_{s}\rightarrow\infty$ (indicating
spatial confinement);   the phase transition to a Coulomb phase
disappears  together  with  the layer phase in the infinite volume
limit. We mention also that the spatial-transverse helicity modulus
($h_{s5}(\beta)$) remains zero throughout the transition.
\begin{figure}[h!]
\begin{center}
\includegraphics[width=5.5 cm,height=8.5 cm, angle=270]{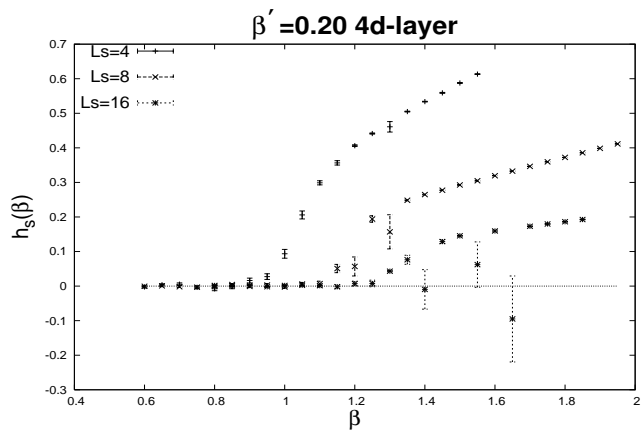}
\end{center}
\caption{The spatial helicity modulus $h_{s}$ is strictly zero for each volume $V=2\times L_{s}^{4}$
until the pseudo-critical value $\beta_{c}(L_{s})$ is approached. For $\beta > \beta_{c}(L_{s})$ the $h_{s}$
tends to zero as $\frac{1}{L_{s}}$ for constant $\beta$.}
\label{4-2-1}
\end{figure}

In Fig.~8(a) we present the temporal helicity modulus $h_{t}$;
also, in Fig.~8(b) we present the temporal-transverse helicity
modulus $h_{t5}$ versus $\beta$, for three different volumes. The
two quantities have the same behavior: both take values equal to
 zero for $\beta \leq 0.85$ and non-zero for $\beta > 0.85$ and
they increase  with the lattice size $L_{s}$, indicating a coulombic
behavior in the temporal direction. We also note that  the Polyakov
loop  in the temporal direction, a result not shown  here, is zero
for $\beta$ smaller than a critical value ( $\beta_{c}\simeq 0.85$)
and tends to one for $\beta > \beta_{c}$. The transition, for the
three quantities $h_{t}$, $h_{t5}$ and $<|P_{t}|>$,  concerning the
strong confining phase ($\beta \leq 0.85$) to the deconfining phase
($\beta > 0.85$) is a continuous one . Although we do not analyze
further the order of this phase transition we may guess that it may
not be  the case of  a first order phase transition.

All of the results obtained so far advocate to the disappearance of
the layer phase at finite temperature. The layer  gives its place to
a phase showing a  confining behavior in the 4d subspaces (formed
 by the three spatial coordinates and the transverse one) and a
coulombic behavior along  the temporal direction.

%\newpage
\begin{figure}[!h]
\begin{center}
\subfigure[\quad]
{\includegraphics[scale=0.3,angle=270]{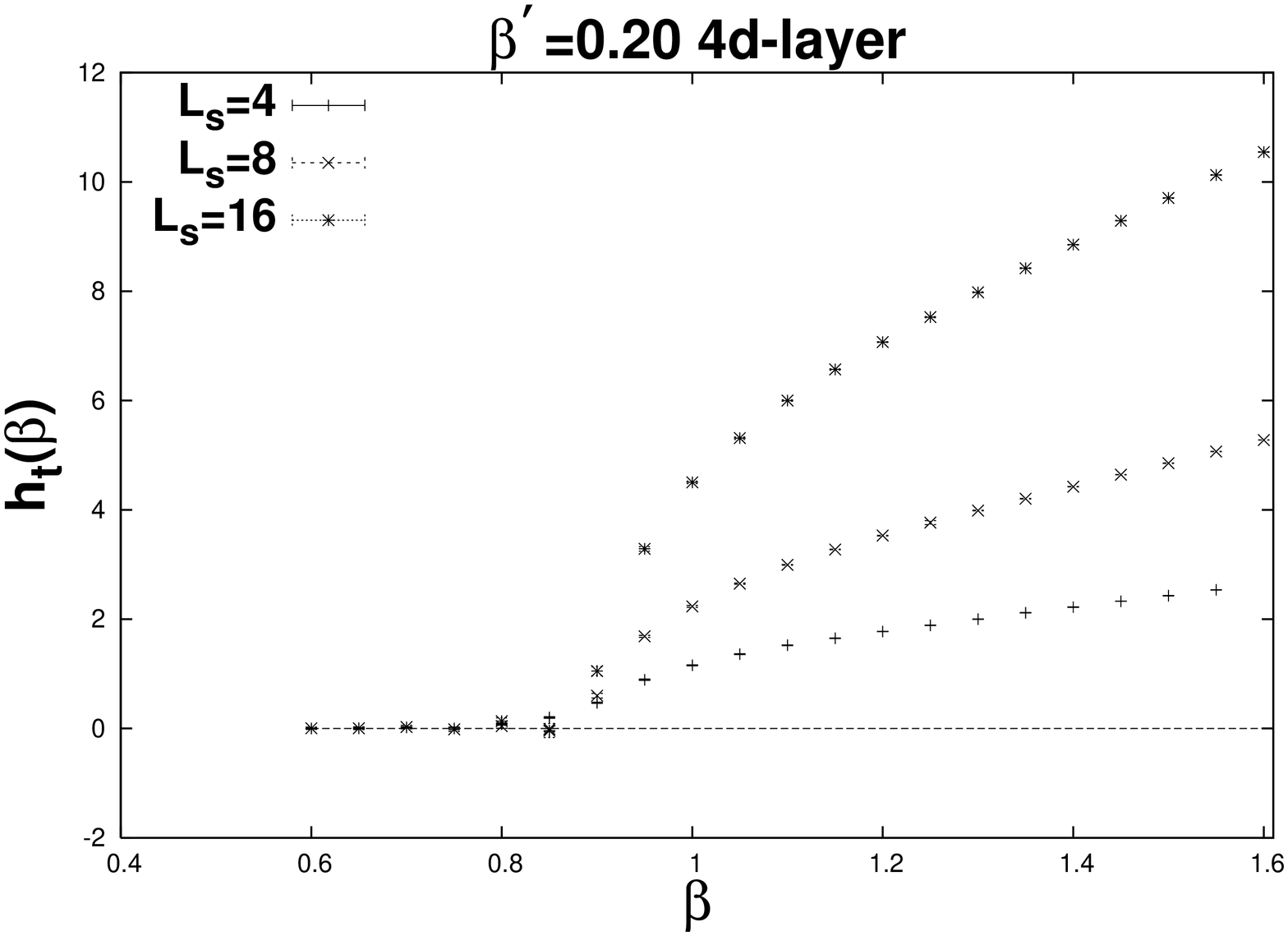}}\quad
\subfigure[\quad]
{\includegraphics[scale=0.3,angle=270]{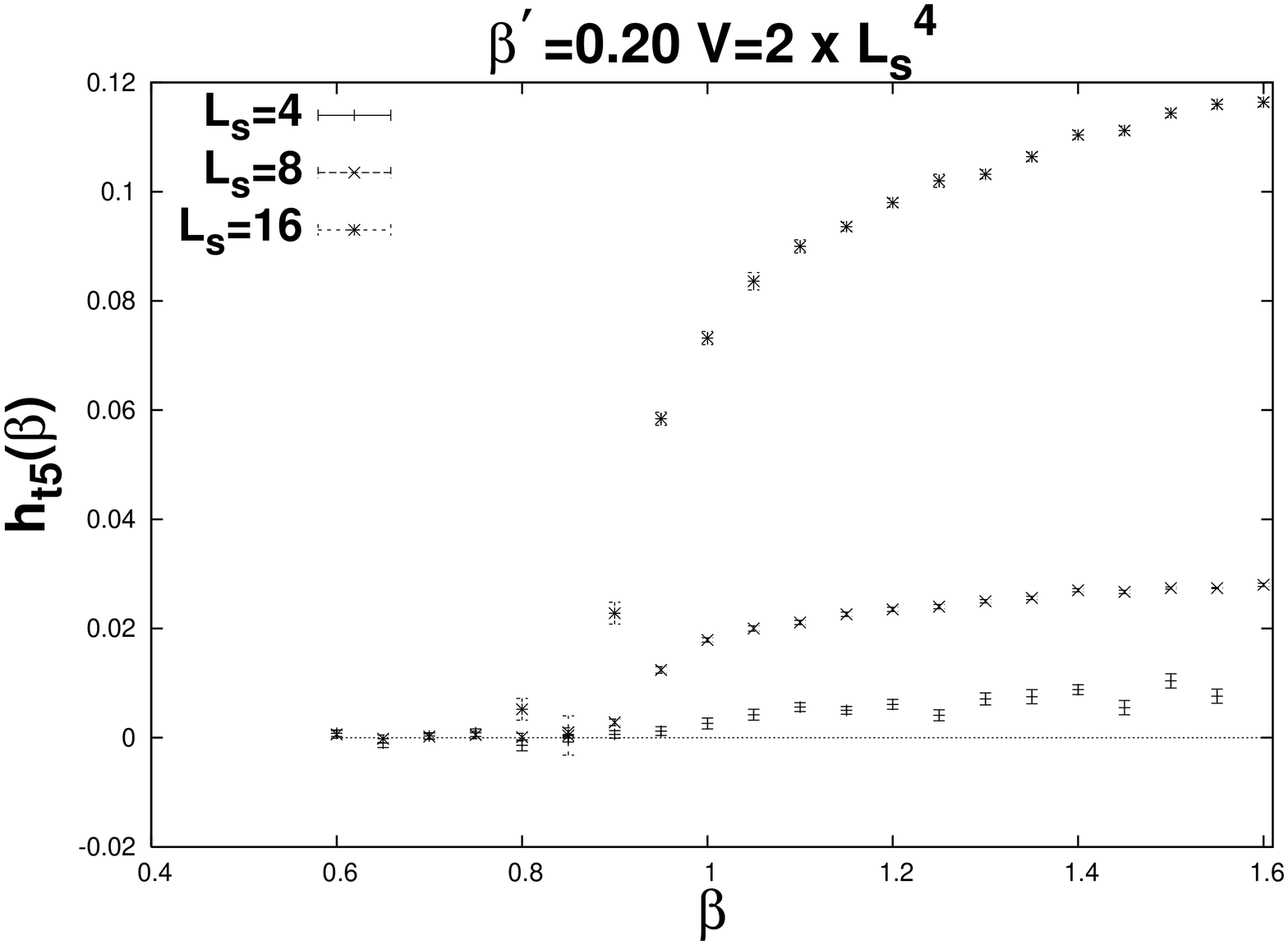}} \caption{The
temporal helicity modulus $h_{t}$ (a) and the temporal-transverse
helicity modulus $h_{t5}$ (b) for $L_{t}=2$ and $\beta^{'}=0.20$
versus $\beta$. The $h_{t}$ is evaluated on the 4d-subspaces
($L_{t}\times L_{s}^{3}$) and scales as $L_{s}$ for
$\beta>\beta_{c}$. The $h_{t5}$ is evaluated on the whole lattice
and scales as $L_{s}^{2}$ for $\beta>\beta_{c}$. }
\end{center}
\end{figure}
%\newpage
\subsection{Moving  along the line ${\mathbf{\beta =1.10}}$}
As we have seen in the previous sections the system undergoes a
continuous phase transition from the strong, confining phase, to a
new phase. The transition point for $\beta^{'}=0$ is  shown to be
$\beta_{c}\simeq 0.90$ and for $\beta^{'}=0.20$ it is slightly
smaller being in  the interval $0.85\leq \beta_{c}<0.90$ region. In
order to study the nature and the extent of the new phase, we choose
to keep $\beta$ fixed at the value of 1.10 and let $\beta^{'}$ to
vary. In Figs.~9(a) and 9(b) we present the spatial helicity
modulus $h_{s}(\beta^{'})$ and the spatial-transverse helicity
modulus $h_{s5}(\beta^{'})$ for three values of the volume. The
$h_{s}$ and $h_{s5}$ are zero,  within the statistical error, for
$\beta^{'}$ smaller than 0.445 signaling disordering in the spatial
and transverse directions. This phase is the continuation of the
$\beta^{'}=0$ phase to non zero values of $\beta^{'}$. The 3d U(1)
theory obtained through dimensional reduction for $\beta^{'}=0$,  is
extended (for $0\leq \beta^{'} \leq 0.445$) to a 4d dimensionally
reduced U(1) theory in the confining phase. We observe that the
layer phase, consisting of a combination of 4d Coulomb phase and
confinement in the extra dimension, becomes a deconfined phase.

There is a critical region  defined in the interval  ($0.445\leq
\beta^{'}\leq 0.450$) in which a finite discontinuity in both
quantities ($h_{s},h_{s5}$) is shown up . For $\beta^{'} >
\beta^{'}_{c}$ the spatial helicity modulus is non zero and almost
constant which is a  characteristic of a Coulomb phase. The value of
$h_{s5}(\beta^{'})$ increases linearly with $\beta^{'}$, following
the lattice weak coupling expansion, approaching $h_{s}(\beta^{'})$
as $\beta^{'}\rightarrow \beta$. The values of $h_{s}$ and $h_{s5}$
in Fig.~9 are divided by $L_{t}$ and are independent of the
spatial lattice size $L_{s}$. The spatial helicity modulus gives the
renormalized coupling $\beta_{R}=\frac{1}{e^{2}_{R}}$ of the 5D U(1)
theory in the Coulomb phase which is fixed by the value of
$\beta=1.10$\cite{FV}.

The temporal and the temporal-transverse helicity modulus  ( not
shown here), remain  non zero and increase  with $\beta^{'}$. Also
the temporal Polyakov loop is non zero which is a  signal of a
finite temperature phase.
\begin{figure}[!h]
\begin{center}
\subfigure[\qquad]
{\includegraphics[scale=0.30,angle=270]{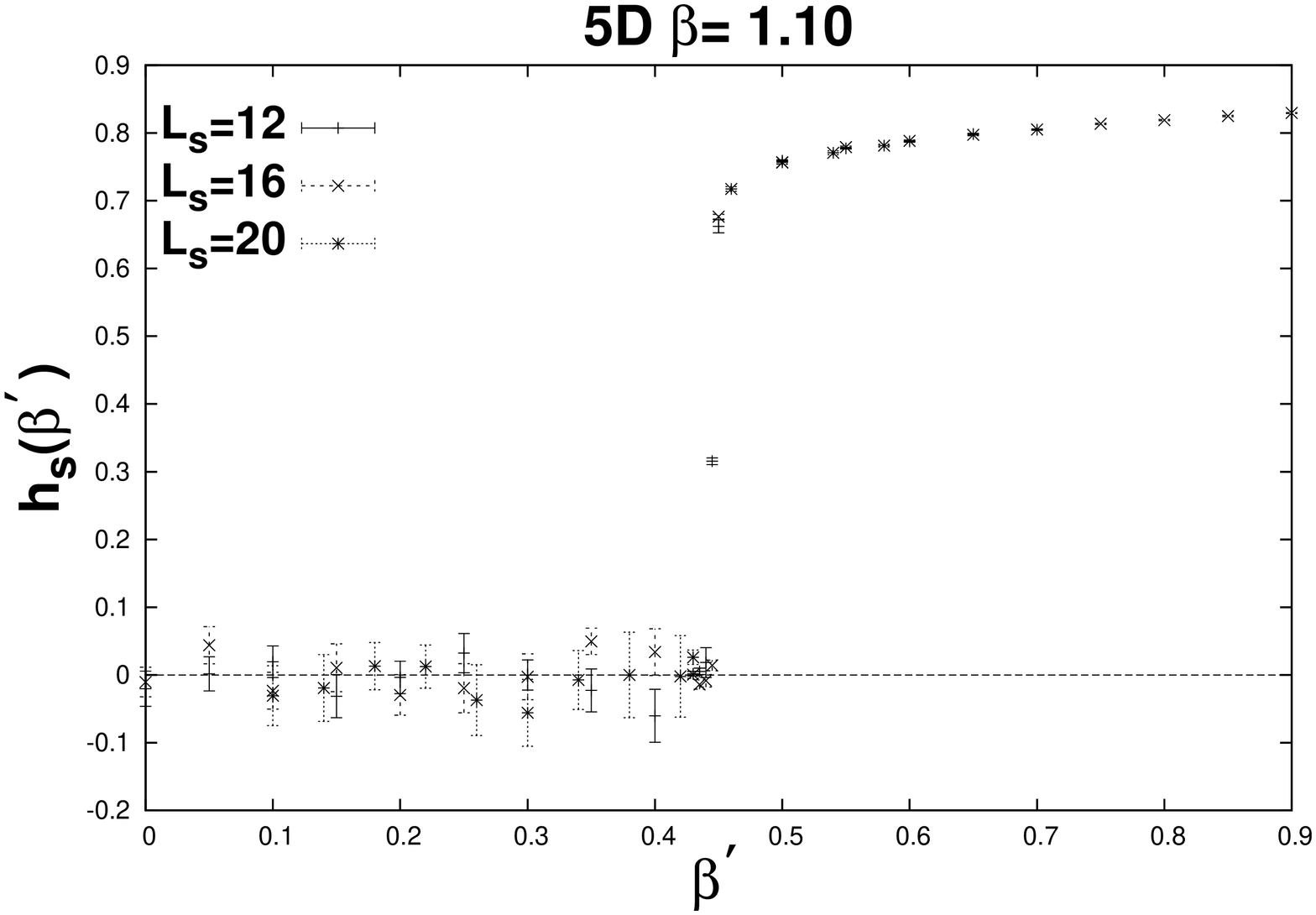}}\quad
\subfigure[\qquad]
{\includegraphics[scale=0.30,angle=270]{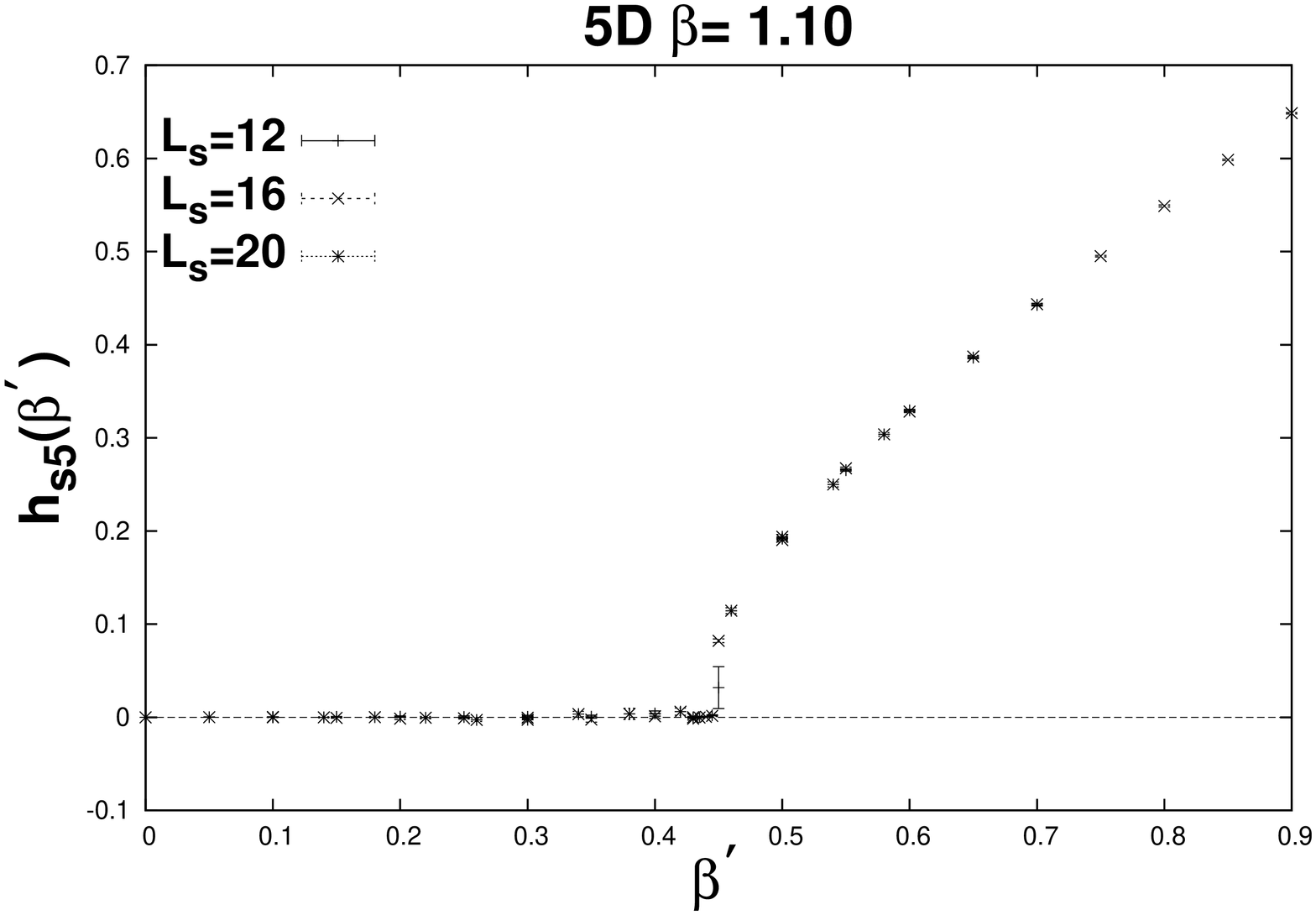}} \caption{The
spatial helicity modulus (a) and the spatial-transverse helicity
modulus (b) as a function of $\beta^{'}$, measured for the temporal
lattice size $L_{t}=2$. The critical value of $\beta^{'}$ remains
constant  with the lattice volume.}
\end{center}
\end{figure}

By close inspection of Fig.~4 ($\beta^{'}=0$) and Fig.~7
($\beta^{'}=0.20$), it becomes obvious  that for a constant value of
$\beta$ the spatial helicity modulus is non-zero for some of the
volumes that we used and it vanishes as  the spatial volume
increases beyond a definite  value. For  $\beta=1.10$, for example,
 the lattice size $L_{s}=16$ is enough to show the correct
thermodynamic limit behavior. If we move to larger values of
$\beta$, like $\beta=1.40$, we have to use a spatial size of the
order $L_{s}\geq 24$ in order  to find the correct behavior. This is
beyond our current computer capabilities.
%\newpage

In Fig.~10 we sketch, roughly, the phase diagram for $L_{t}=2$ in
the ($\beta,\beta^{'}$) plane. There are three phases with different
behavior of the observables we used:
\begin{enumerate}
\item  5D  confining phase with: $P_{t}=0$, $h_{t}=0$, $h_{t5}=h_{s5}=0$ and $h_{s}=0$
\item  Finite temperature 5D Coulomb phase: $P_{t}\neq 0$,\quad $h_{t},h_{s},h_{t5}$ and $h_{s5}\neq0$
\item Dimensionally reduced 4d confining phase-temporal Coulomb
$P_{t}\neq 0$,\quad $h_{t}\neq 0$ ,$h_{t5}\neq0$ and $h_{s}, h_{s5}=0$
\end{enumerate}
%\newpage
\begin{figure}[hb!]
\begin{center}
\includegraphics[width=6 cm,height=9 cm, angle=270]{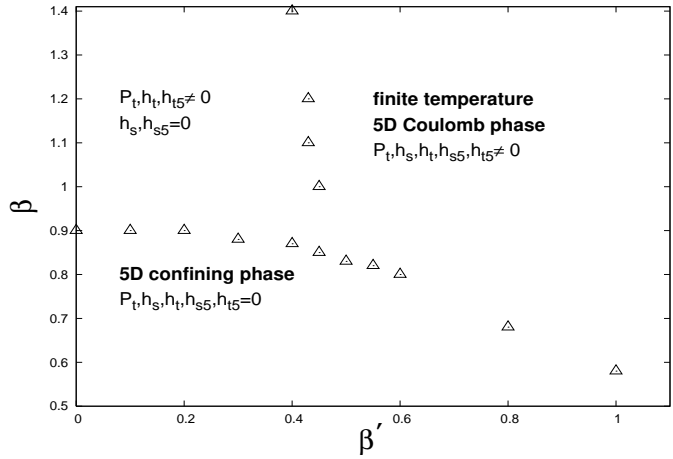}
\end{center}
\caption{A rough sketch for the phase diagram of the model for
$L_{t}=2$. There are three different phases, a 5D confining phase, a
5D Coulomb phase in finite temperature and a new one characterized
as temporal Coulomb-4d confining.} \label{4-2-2}
\end{figure}

From the discussion in section 3.2 for $\beta^{'}=0$ we argue that
the critical temperature for the appearance of the phase diagram of
Fig.~10 it is the zero temperature. The reason is that the layer
phase strongly depends on the existence of the phase transition in
the Coulomb phase for $\beta^{'}=0$. All the results we have
presented in section 3.2 for $T > 0$ and $\beta^{'}=0$ point to a 3d
confining phase, in the infinite volume limit, for $\beta$ larger
than a critical value $\beta_{c}(L_t)$. A Coulomb phase does not
seem to be the case. From this analysis we conclude that the phase
diagram presented in Fig.~10 is reproduced for every temperature
bigger than zero. Especially for $\beta>\beta_{c}(L_t)$ and
$0<\beta^{'}<\beta^{'}_{c}(L_t)$ we have a 4d confining-temporal
coulombic phase instead of a layer phase. Two charges are not
anymore localized (confined) on a three dimensional subspace (brane)
but the temperature gives the possibility of having interactions
between the neighbor three dimensional subspaces. It seems that
there are two characteristic correlation lengths in this deconfining
phase. The correlation length given by the spatial string tension
and a second one characterising the thickness of the brane given by
the interaction in the transverse direction and the temperature. We
did not study quantitatively these two correlation lengths at finite
temperature but we may  easily see  from the Fig.~6 what happens in
two limiting cases. For $L_{t}=1$ (infinite temperature)  we have a
4d U(1) gauge theory in the strong confining phase and the two
correlation lengths are indistinguishable and  approach  each other.
For the zero temperature case on the other hand,  there is no
spatial string tension;  we get a massless photon on the branes.
Note that in this case the branes  are characterized by zero
thickness. In between these two limiting cases we expect a
continuous change in the behavior depending strongly  on  the
temperature.

%\newpage
\section{Discussion}
The extra dimensional models, like the brane models, are well
studied mainly in the zero temperature case. But if we imagine that
our brane world is  a part of the Universe history then a study of
the brane models at  high temperature is required. In this paper
 we tried  to do a first approach to this open problem,
namely the behavior of brane models in high temperature  (though
neglecting  the gravity effects). We believe that our toy model of
five dimensional U(1) anisotropic lattice gauge theory has all the
required essential characteristics. This model has a very rich phase
diagram with respect to the temperature (see the discussion  in
Sections 3 and 4) summarized in Figures 6 and 10.

Concluding we could note that the layer phase for zero temperature
(with a massless photon on the brane and confinement in the extra
dimensions) gives its place to a deconfined phase at non-zero
temperature. In this phase the three spatial dimensions and the
transverse one form  a 4d subspace with confining properties, while
 the temporal direction shows a coulombic behavior.

\section{Acknowledgments}
We would like to thank K. Anagnostopoulos, P. Dimopoulos and G.
Koutsoumbas for their help and support with the  manuscript. In
particular  we wish to thank K.Anagnostopoulos for his help
concerning  the 4D $\bold{XY}$ model. We are also  grateful to P. de
Forcrand for his comments and  discussions on this work.


\begin{thebibliography}{99}
\bibitem{FN}
  Y.~K.~Fu and H.~B.~Nielsen,
  %``A Layer Phase In A Nonisotropic U(1) Lattice Gauge Theory: Dimensional
  %Reduction A New Way,''
  Nucl.\ Phys.\  B {\bf 236} (1984) 167.
\bibitem{RS}
  L.~Randall and R.~Sundrum,
  %``An alternative to compactification,''
  Phys.\ Rev.\ Lett.\  {\bf 83}, 4690 (1999)
  [arXiv:hep-th/9906064];\\
  L.~Randall and R.~Sundrum,
  %``A large mass hierarchy from a small extra dimension,''
  Phys.\ Rev.\ Lett.\  {\bf 83}, 3370 (1999)
  [arXiv:hep-ph/9905221].
\bibitem{DFKK}
  P.~Dimopoulos, K.~Farakos, A.~Kehagias and G.~Koutsoumbas,
  %``Lattice evidence for gauge field localization on a brane,''
  Nucl.\ Phys.\  B {\bf 617}, 237 (2001)
  [arXiv:hep-th/0007079].
\bibitem{FV}
K.~Farakos and S.~Vrentzos,
  %``Establishment of the Coulomb law in the layer phase of a pure U(1) lattice
  %gauge theory,''
  Phys.\ Rev.\  D {\bf 77}, 094511 (2008)
  [arXiv:0801.3722 [hep-lat]].
\bibitem{DvS}
  G.~R.~Dvali and M.~A.~Shifman,
  %``Domain walls in strongly coupled theories,''
  Phys.\ Lett.\  B {\bf 396} (1997) 64
  [Erratum-ibid.\  B {\bf 407} (1997) 452]
  [arXiv:hep-th/9612128].
\bibitem{FP}
  K.~Farakos and P.~Pasipoularides,
  %``Gravity-induced instability and gauge field localization,''
  Phys.\ Lett.\  B {\bf 621}, 224 (2005)
  [arXiv:hep-th/0504014];
  K.~Farakos and P.~Pasipoularides,
  %``RS2-brane world scenario with a nonminimally coupled bulk scalar field,''
  Phys.\ Rev.\  D {\bf 73}, 084012 (2006)
  [arXiv:hep-th/0602200];
  K.~Farakos and P.~Pasipoularides,
  %``Gauss-Bonnet gravity, brane world models, and non-minimal coupling,''
  Phys.\ Rev.\  D {\bf 75}, 024018 (2007)
  [arXiv:hep-th/0610010].
  P.~Pasipoularides and K.~Farakos,
  %``Brane World Scenario In The Presence Of A Non-Minimally Coupled Bulk Scalar
  %Field,''
  J.\ Phys.\ Conf.\ Ser.\  {\bf 68} (2007) 012041.
\bibitem{HCFu}
L.~X.~Huang, T.~L.~Chen and Y.~K.~Fu,
  %``Is a layer phase dissolved in the early universe,''
  Phys.\ Lett.\  B {\bf 329}, 175 (1994).
\bibitem{DFKKN}
 P.~Dimopoulos, K.~Farakos, C.~P.~Korthals-Altes, G.~Koutsoumbas and S.~Nicolis,
  %``Phase structure of the 5D Abelian Higgs model with anisotropic
  %couplings,''
  JHEP {\bf 0102}, 005 (2001)
  [arXiv:hep-lat/0012028].
\bibitem{AnisoH}
P.~Dimopoulos and K.~Farakos,
  %``Layered Higgs phase as a possible field localisation on a brane,''
  Phys.\ Rev.\  D {\bf 70} (2004) 045005
  [arXiv:hep-ph/0404288];
  P.~Dimopoulos, K.~Farakos and G.~Koutsoumbas,
  %``The phase diagram for the anisotropic SU(2) adjoint Higgs model in  5D:
  %Lattice evidence for layered structure,''
  Phys.\ Rev.\  D {\bf 65} (2002) 074505
  [arXiv:hep-lat/0111047];
  P.~Dimopoulos, K.~Farakos and S.~Nicolis,
  %``Multi-layer structure in the strongly coupled 5D Abelian Higgs model,''
  Eur.\ Phys.\ J.\  C {\bf 24} (2002) 287
  [arXiv:hep-lat/0105014].
\bibitem{DeGT}
T.~A.~DeGrand and D.~Toussaint,
  %``Topological Excitations And Monte Carlo Simulation Of Abelian Gauge
  %Theory,''
  Phys.\ Rev.\  D {\bf 22}, 2478 (1980).
\bibitem{CIS}
M.~N.~Chernodub, E.~M.~Ilgenfritz and A.~Schiller,
  %``A lattice study of 3-D compact QED at finite temperature,''
  Phys.\ Rev.\  D {\bf 64}, 054507 (2001)
  [arXiv:hep-lat/0105021];
  M.~N.~Chernodub, E.~M.~Ilgenfritz and A.~Schiller,
  %``Monopoles, confinement and deconfinement in lattice compact QED in  (2+1)D
  %with external fields,''
  Nucl.\ Phys.\ Proc.\ Suppl.\  {\bf 106}, 703 (2002)
  [arXiv:hep-lat/0110038].
\bibitem{Berg}
B.~A.~Berg and A.~Bazavov,
  %``Non-perturbative U(1) gauge theory at finite temperature,''
  Phys.\ Rev.\  D {\bf 74}, 094502 (2006)
  [arXiv:hep-lat/0605019];
  B.~A.~Berg and A.~Bazavov,
  %``Critical exponents for U(1) lattice gauge theory at finite temperature,''
  PoS {\bf LAT2006}, 061 (2006)
  [arXiv:hep-lat/0609006].
\bibitem{FdV}
M.~Vettorazzo and P.~de Forcrand,
  %``Electromagnetic fluxes, monopoles, and the order of the 4d compact U(1)
  %phase transition,''
  Nucl.\ Phys.\  B {\bf 686}, 85 (2004)
  [arXiv:hep-lat/0311006];
  M.~Vettorazzo and P.~de Forcrand,
  %``The helicity modulus in gauge field theories,''
  Nucl.\ Phys.\ Proc.\ Suppl.\  {\bf 129}, 739 (2004)
  [arXiv:hep-lat/0311007];
  M.~Vettorazzo and P.~de Forcrand,
  %``Finite temperature phase transition in the 4d compact U(1) lattice  gauge
  %theory,''
  Phys.\ Lett.\  B {\bf 604}, 82 (2004)
  [arXiv:hep-lat/0409135].
\bibitem{Borgs}
  C.~Borgs,
  %``Area Law For Spatial Wilson Loops In High Temperature Lattice Gauge
  %Theories,''
  Nucl.\ Phys.\  B {\bf 261} (1985) 455.
\bibitem{DFV}
P.~Dimopoulos, K.~Farakos and S.~Vrentzos,
  %``The 4-D layer phase as a gauge field localization: Extensive study of  the
  %5-D anisotropic U(1) gauge model on the lattice,''
  Phys.\ Rev.\  D {\bf 74}, 094506 (2006)
  [arXiv:hep-lat/0607033].
\bibitem{KAN}
A.~Hulsebos, C.~P.~Korthals-Altes and S.~Nicolis,
  %``Gauge theories with a layered phase,''
  Nucl.\ Phys.\  B {\bf 450}, 437 (1995)
  [arXiv:hep-th/9406003];
  C.~P.~Korthals-Altes, S.~Nicolis and J.~Prades,
  %``Chiral defect fermions and the layered phase,''
  Phys.\ Lett.\  B {\bf 316}, 339 (1993)
  [arXiv:hep-lat/9306017].
\bibitem{KT}
  A.~Kehagias and K.~Tamvakis,
  %``Localized gravitons, gauge bosons and chiral fermions in smooth spaces
  %generated by a bounce,''
  Phys.\ Lett.\  B {\bf 504} (2001) 38
  [arXiv:hep-th/0010112].
\bibitem{Cardy}
J.~L.~Cardy,
  %``Universal Properties Of U(1) Gauge Theories,''
  Nucl.\ Phys.\  B {\bf 170}, 369 (1980);
  G.~'t Hooft,
  %``A Property Of Electric And Magnetic Flux In Nonabelian Gauge Theories,''
  Nucl.\ Phys.\  B {\bf 153}, 141 (1979;
  J.~Groeneveld, J.~Jurkiewicz and C.~P.~Korthals Altes,
  %``Local Order Parameter In Twisted Gauge Fields,''
  Phys.\ Lett.\  B {\bf 92}, 312 (1980).
\bibitem{GH}
  A.~P.~Gottlob and M.~Hasenbusch,
  %``Critical behavior of the 3-D XY model: A Monte Carlo study,''
  Physica A {\bf 201} (1993) 593.
\bibitem{Arnold}
G.~Arnold, B.~Bunk, T.~Lippert and K.~Schilling,
  %``Compact QED under scrutiny: It's first order,''
  Nucl.\ Phys.\ Proc.\ Suppl.\  {\bf 119}, 864 (2003)
  [arXiv:hep-lat/0210010];
  G.~Arnold, T.~Lippert, K.~Schilling and T.~Neuhaus,
  %``Finite size scaling analysis of compact QED,''
  Nucl.\ Phys.\ Proc.\ Suppl.\  {\bf 94}, 651 (2001)
  [arXiv:hep-lat/0011058].

\end{thebibliography}
\end{document}